\documentclass[journal]{IEEEtran}
\usepackage{cite}

\ifCLASSINFOpdf
\else
\fi

\usepackage{xcolor}
\usepackage{graphicx}
\usepackage{amsmath, bbm}
\usepackage{amssymb}
\usepackage{tabularx, booktabs}
\usepackage{multirow}
\usepackage{array,multirow}
\usepackage{cite}
\usepackage{textcomp}
\usepackage{footnote}
\usepackage{fancyhdr}

\usepackage{algorithm,algorithmic}

\newcommand{\norm}[1]{\left\lVert#1\right\rVert}

\begin{document}
%
\title{Affine Transformation-Based Deep Frame Prediction }

\author{Hyomin~Choi,~\IEEEmembership{Student Member,~IEEE,}
        and~Ivan~V.~Baji\'{c},~\IEEEmembership{Senior Member,~IEEE}
\thanks{H. Choi and I. V. Baji\'{c} are with the School of Engineering Science, Simon Fraser University, BC, V5A 1S6, Canada. E-mail: chyomin@sfu.ca, ibajic@ensc.sfu.ca}}

\markboth{Accepted for publication}
{Shell \MakeLowercase{\textit{et al.}}: Bare Demo of IEEEtran.cls for IEEE Journals}

\maketitle

\thispagestyle{empty}
\renewcommand{\headrulewidth}{0.0pt}
\thispagestyle{fancy}
\lhead{}
\chead{Copyright \copyright 2021 IEEE. Personal use of this material is permitted. However, permission to use this material for any other purposes must be obtained from the IEEE by sending an email to pubs-permissions@ieee.org.}
\rhead{}
\lfoot{}
\cfoot{}
\rfoot{}

\begin{abstract}
We propose a neural network model to estimate the current frame from two reference frames, using affine transformation and adaptive spatially-varying filters. 
The estimated affine transformation allows for using shorter filters compared to existing approaches for deep frame prediction. 
The predicted frame is used as a reference for coding the current frame. Since the proposed model is available at both encoder and decoder, there is no need to code or transmit motion information for 
the predicted frame. By making use of dilated convolutions and reduced filter length, our model is significantly smaller, yet more accurate, than any of the neural networks in prior works on this topic. Two versions of the proposed model -- one for uni-directional, and one for bi-directional prediction -- are trained using a combination of Discrete Cosine Transform (DCT)-based $\ell_1$-loss with various transform sizes, multi-scale Mean Squared Error (MSE) loss, and an object context reconstruction loss. 
The trained models are integrated with the HEVC video coding pipeline. 
The experiments show that the proposed models achieve about 7.3\%, 5.4\%, and 4.2\% bit savings for the luminance component on average in the Low delay P, Low delay, and Random access configurations, respectively.

\end{abstract}


\IEEEpeerreviewmaketitle

\section{Introduction}
\IEEEPARstart{N}{eural} network-aided prediction tools have recently become a topic of interest in image and video coding research. The latest studies~\cite{dnn_aided_video_coding, dnn_video_coding_tcsvt} show that new coding tools built upon deep neural networks (DNNs) are able to achieve superior rate-distortion (RD) performance compared to the conventional coding tools. Due to such promising results, the standardization community also keeps track of this technological trend and considers a possible inclusion of these tools into 
future standards~\cite{jvet_j1003, ahg9_macao}. 

In a typical video coding pipeline, there are several components~\cite{jvet_k0222, JVET_J0037_sandiego_intra_hhi, he2018enhancing, zhang2017learning, yan2018convolutional, huo2018convolutional, zhao2018cnn, zhao2018enhanced, zhao2019enhanced, hevc_with_sep_conv_for_ra, deep_frame_prediction, xia2019deep, lee2020deep, extrapolation_with_reference_alignment} and optimization processes~\cite{xu2018reducing, wang2018fast, xu2017cnn}  that can be aided or replaced completely by DNNs. In particular, many studies regarding DNN-aided inter prediction 
have been published in the last few years~\cite{zhang2017learning, yan2018convolutional, huo2018convolutional, zhao2018cnn, zhao2018enhanced, zhao2019enhanced, hevc_with_sep_conv_for_ra, deep_frame_prediction, xia2019deep, lee2020deep, extrapolation_with_reference_alignment}. In order to minimize residuals after inter prediction,~\cite{zhang2017learning, yan2018convolutional} proposed networks that learn interpolation filters, while~\cite{huo2018convolutional, zhao2018cnn, zhao2018enhanced} developed refinement networks using motion compensated patches as input. In~\cite{extrapolation_with_reference_alignment}, a lighter model for uni-directional prediction (extrapolation) was proposed. Models presented in~\cite{hevc_with_sep_conv_for_ra, deep_frame_prediction, xia2019deep, lee2020deep} share a common network architecture originally developed in~\cite{Niklaus_ICCV_2017} for frame rate up-conversion (FRUC). This network estimates a halfway frame between two input frames (i.e, bi-directional prediction, or interpolation), which can be used as a prediction of the current frame in video coding. The network learns to compute content-adaptive spatially-varying filter kernels, 
which are used to generate the predicted frame. 

\begin{figure}[t]
    \centering
    \begin{minipage}[b]{0.46\linewidth}
    \centering
    \includegraphics[width=\textwidth]{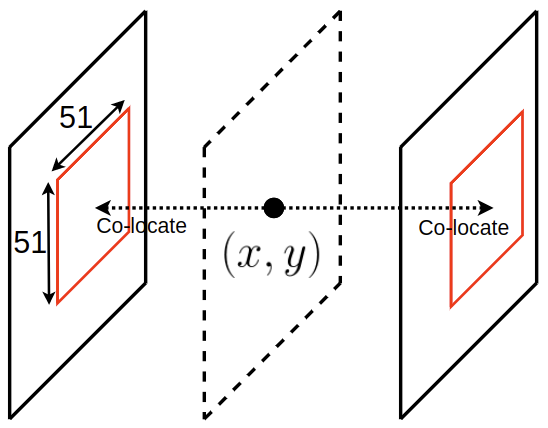}
    \centerline{(a) Previous works~\cite{hevc_with_sep_conv_for_ra, deep_frame_prediction, xia2019deep, lee2020deep} }\medskip
    \end{minipage}
    \hspace{0.04\linewidth}
    \centering
    \begin{minipage}[b]{0.46\linewidth}
    \centering
    \includegraphics[width=\textwidth]{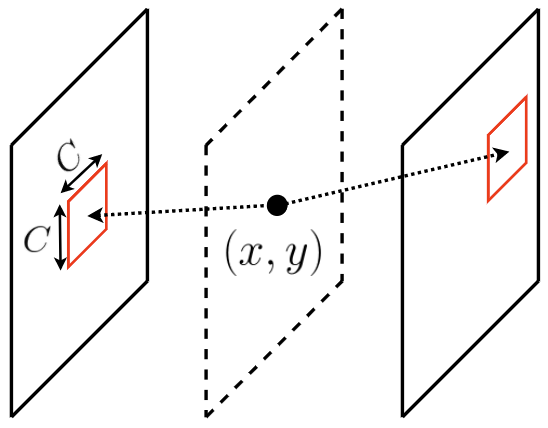}
    \centerline{(b) Proposed method}\medskip
    \end{minipage}
    
    \centering
    \begin{minipage}[b]{0.485\linewidth}
    \centering
    \includegraphics[width=\textwidth]{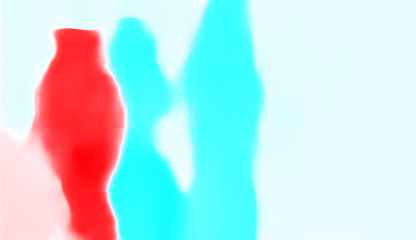}
    \end{minipage}
    \hspace{0.005\linewidth}
    \centering
    \begin{minipage}[b]{0.485\linewidth}
    \centering
    \includegraphics[width=\textwidth]{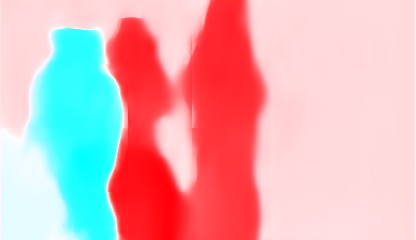}
    \end{minipage}
    \centerline{(c) Visualized motion fields based on learned affine parameters }\medskip
    
    \centering
    \begin{minipage}[b]{0.485\linewidth}
    \centering
    \includegraphics[width=\textwidth]{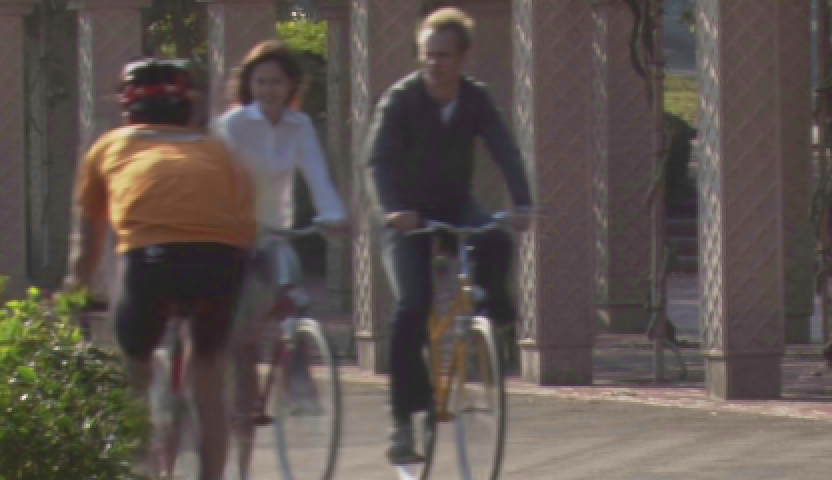}
    \centerline{(d) Estimated output frame}
    \end{minipage}
    \hspace{0.005\linewidth}
    \centering
    \begin{minipage}[b]{0.485\linewidth}
    \centering
    \includegraphics[width=\textwidth]{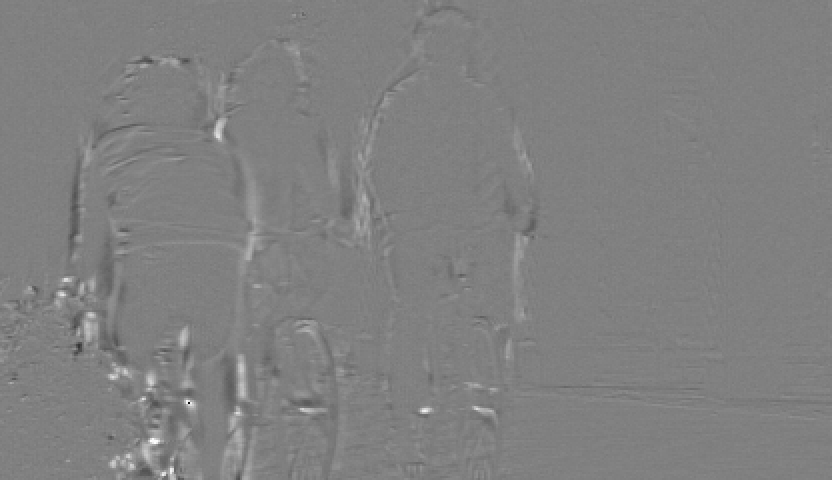}
    \centerline{(e) Visualized residue}
    \end{minipage}
    
\caption{Comparison of two approaches for bi-directional prediction and an example of an affine-based motion field.}
\label{fig:comp_bi-prediction}
\end{figure}

Fig.~\ref{fig:comp_bi-prediction}(a) illustrates bi-directional prediction using the computed filter kernels in~\cite{hevc_with_sep_conv_for_ra, deep_frame_prediction, xia2019deep, lee2020deep}. The network generates a pair of 1-D adaptive separable filters (ASFs) with the length of $51$, whose outer product defines a $51\times51$ 2-D filter kernel. This kernel is applied to collocated areas 
in the reference frames. In this setup, having a relatively large filter kernel seems crucial in order to cover a wide range of possible motions. In contrast,  conventional motion estimation and compensation~\cite{hevc_sze} uses much smaller filters with constant coefficients, but aided with motion vectors, which are transferred to the decoder. In fact, the results in~\cite{deep_frame_prediction} showed that conventional motion estimation and compensation in HEVC produces more accurate frame prediction (in terms of PSNR) than the DNN from~\cite{Niklaus_ICCV_2017}. Nonetheless, the DNN-based frame prediction methods~\cite{hevc_with_sep_conv_for_ra, deep_frame_prediction, xia2019deep, lee2020deep} achieve better rate-distortion performance than HEVC mostly because they avoid the transfer of motion information.  

In this paper, we propose a novel neural network architecture that combines the advantages of DNN-based and conventional motion estimation. Specifically, the proposed network estimates an affine transformation (translation and scale) locally, and then uses this information to position the learned spatially varying filters in order to perform frame prediction, as illustrated in Fig.~\ref{fig:comp_bi-prediction}(b). As a result, filters can be much smaller (e.g., $C=8$) and can be estimated using a lighter network than in~\cite{hevc_with_sep_conv_for_ra, deep_frame_prediction, xia2019deep, lee2020deep}. Still, no motion information needs to be transferred to the decoder, which leads to superior rate-distortion performance. 
For example, Fig.~\ref{fig:comp_bi-prediction}(c) shows two estimated motion fields: 
backward (left) and forward (right) flow from the middle of the two inputs in time,
based on the learned affine parameters, where hue and saturation indicate orientation and magnitude of motion vectors, respectively. It can be seen that the motion discontinuity roughly corresponds to object boundaries, indicating that the general motion trend has been well-captured by the affine parameters. The role of learned filters in this context is to refine this motion information and improve frame prediction accuracy.
For example, Fig.~\ref{fig:comp_bi-prediction}(d) shows the DNN-generated output frame by our method using the learned motion with input reference frames, and (e) shows 
the visualized residual between the output frame and the corresponding ground truth frame by adding 128.

The paper is organized as follows. In Section~\ref{sec:literature_review}, relevant prior work is reviewed in more detail and the contributions of this paper are summarized. Section~\ref{sec:proposed_method} presents the proposed network architecture, loss functions, and training strategy. Integration of the proposed network into the HEVC video coding pipeline is described in Section~\ref{sec:network_strategies}, followed by experiments 
in Section~\ref{sec:experimental_results}. Finally, the paper is concluded in Section~\ref{sec:conclusion}.

\section{Prior work and our contribution}
\label{sec:literature_review}
Most relevant prior studies~\cite{zhao2019enhanced, hevc_with_sep_conv_for_ra, deep_frame_prediction, xia2019deep, lee2020deep} make use of the frame rate up-conversion (FRUC) network developed by~\cite{Niklaus_ICCV_2017, liu2017video} to generate predicted 
frame halfway between two input frames. The network  in~\cite{Niklaus_ICCV_2017} is comprised of a U-Net~\cite{Ronneberger2015UNetCN}, which extracts features from input frames, and a kernel estimation module that computes pairs of 1-D  adaptive separable filters (ASFs) based on the extracted features. Spatially-varying 2-D filters 
are computed as outer products of 1-D ASFs. Zhao \textit{et al.}~\cite{hevc_with_sep_conv_for_ra} incorporated this FRUC network, with pre-trained weights, into HEVC. The FRUC network takes two decoded frames and generates a bi-predicted frame, which is used by largest coding blocks as a reference for B-frame coding. 
A subsequent study by the same group~\cite{zhao2019enhanced} introduced a refinement network to 
improve the output of the FRUC network. 
Furthermore, smaller coding blocks were allowed to use the predicted frame as a reference, to improve coding efficiency. 

In order to deploy these DNN-based prediction approaches in both P and B frames for general compression scenarios, our earlier study~\cite{deep_frame_prediction} proposed a modified architecture of the network~\cite{Niklaus_ICCV_2017}, whose input comprises not only two decoded frames, but also their temporal indices, 
so that the network is explicitly instructed whether to perform extrapolation or interpolation. Huo \textit{et al.}~\cite{extrapolation_with_reference_alignment} focused on frame extrapolation by applying the frame alignment pre-processing between reference frames before using them as inputs to a lightweight prediction network. 
Recently,\cite{xia2019deep, lee2020deep} introduced multi-scale ASFs to the FRUC network~\cite{Niklaus_ICCV_2017}. In~\cite{lee2020deep}, two FRUC networks are employed to estimate halfway frames at different scales, which are then merged by weighted average. In~\cite{xia2019deep}, each set of ASFs is generated individually at three depths in the U-Net subnetwork and then  applied to the properly scaled input frames with  bilinear interpolation to match the resolution of each depth. However, this modification is only required during training. For testing, only the set of ASFs at the first depth is applied to the given input frames to create the intermediate frame. Moreover, for training, Sum of Absolute Transformed Difference (SATD) using Hadamard transform with size of $8\times8$ is used as a loss function to improve the prediction and coding performance.

Compared with conventional motion estimation and motion compensation (MEMC) used in video coding~\cite{hevc_sze}, DNN-based frame prediction offers greater flexibility in motion modeling. However, applying the computed adaptive filters to the collocated areas in reference frames 
necessitates the use of very large filter kernels to capture the motion between frames. 
This also leads to a large number of trainable parameters in the network. To address these problems, Reda \textit{et al.}~\cite{reda2018sdc} proposed a spatial-displaced convolution network (SDC-Net) that applies smaller ASFs to adaptively-chosen positions, rather than collocated areas, in reference frames. However, a large off-the-shelf optical flow network is used to adaptively choose these positions. SDC-Net was evaluated on immediate future frame prediction (i.e., extrapolation). In our case, we achieve a similar effect of adaptively choosing where to apply ASFs, but with a smaller network that estimates affine motion. Moreover, our model can be used for both interpolation and extrapolation. 

Most recently, Bao \textit{et al.}~\cite{bao2019memc} proposed a model called MEMC-Net to address some of these issues. 
MEMC-Net uses five major subnetworks to interpolate a frame between two input frames. Its filter kernel estimator creates input-adaptive $4\times4$ 2-D filter kernels, and the positions to apply the filters are decided based on an estimated optical flow between input frames. However, due to the fact that the estimated optical flow is between the two input frames, not the desired frame and input frames, some assumptions about the linearity of motion are made in order to refer the motion to the desired intermediate frame. 
To further enhance the prediction quality, the network also employs context and mask estimator for the adaptive warping process and a post-processing at the end. Although the network ends up using small filters, its overall size is still quite large, about 67.2 million parameters.


\begin{figure*}[t]
    \centering
    \begin{minipage}[b]{1\linewidth}
    \centering
    \includegraphics[width=\textwidth]{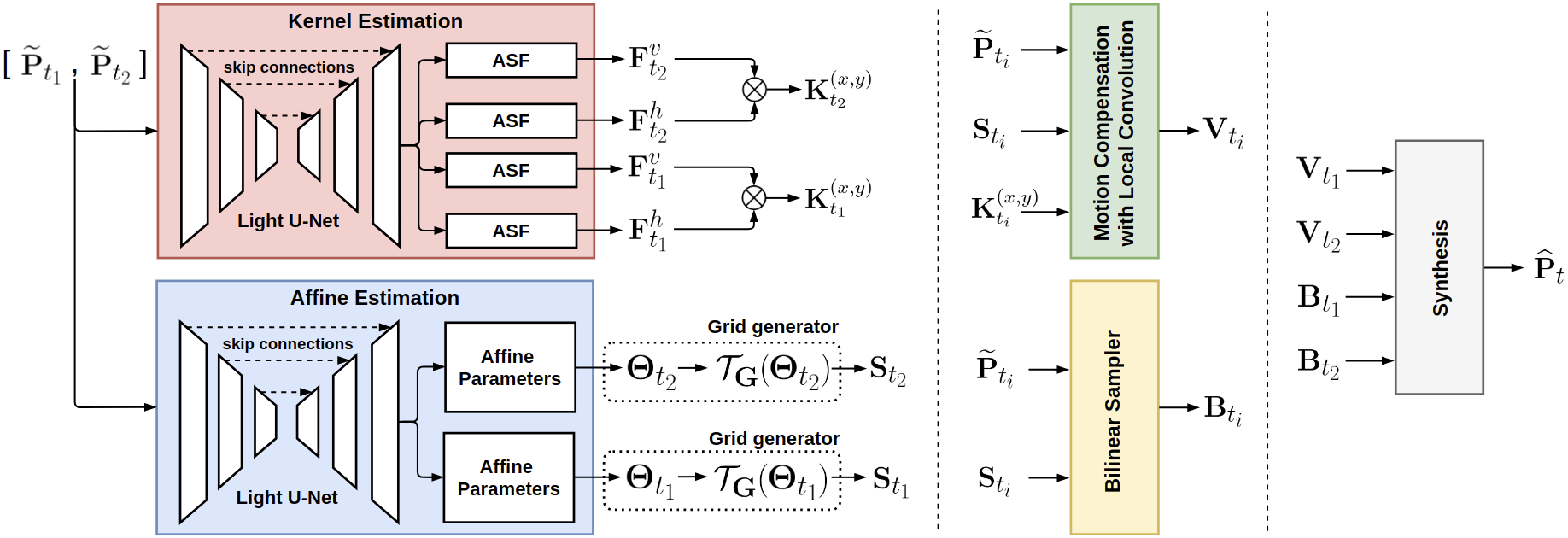}
    \end{minipage}
\caption{Overall architecture of the affine transformation based deep frame prediction. Two patches as input of the kernel and the affine estimation are used to compute spatially-varying filters and affine parameters, respectively. $\otimes$ presents outer product to create spatially-varying 2-D convolution filter kernels for each pixel location. Using the computed filters with affine parameters, four warped samples are generated through the two sampling operations: MCLC and Bilinear Sampler. Finally the output patch is synthesized by GridNet~\cite{niklaus2018context} using the four warped samples as input. Our network is trained and tested in end-to-end manner.}
\label{fig:proposed_architecture}
\end{figure*}

In this work we introduce an affine transformation-based deep frame prediction architecture for video coding. 
Affine motion parameters are estimated using a version of the Spatial Transformer Network~\cite{NIPSjaderberg2015spatial} and mapped to the motion between the desired output frame and the input reference frames, indicating positions where to apply the computed spatially-varying filters. This way, large filter kernels are no longer needed, allowing for potential model size reduction. To actually realize a smaller, yet accurate model for frame prediction,  dilated convolutions~\cite{Yu2016MultiScaleCA} are adopted to cover a large receptive field for motion modeling with fewer parameters. The net result is that  the size of our model is reduced to about 5.5 million parameters, significantly lower compared to any of prior deep model for frame prediction in video coding, while still offering competitive coding performance. 
In summary, the contributions of the present work are as follows:
\begin{itemize}
    \item A novel DNN architecture involving affine motion modeling and computed spatially-varying filters is proposed for P- or B-frame prediction in video coding;
    \item Filter kernels are smaller than in most of the prior work on this topic, and the model is smaller than any of the previously proposed models for frame prediction in video coding, yet it offers competitive coding performance; 
    \item A novel loss function is proposed to encourage coding-friendly motion modeling, with several terms, including a DCT-based $\ell_1$-loss with various transform sizes; 
    \item The network is integrated into the HEVC coding pipeline 
    and compared with prior work. 
\end{itemize}

\section{Proposed DNN for frame prediction}
\label{sec:proposed_method}

Like prior models~\cite{hevc_with_sep_conv_for_ra, zhao2019enhanced, deep_frame_prediction, xia2019deep}, our network requires two previous coded frames from the decoded picture buffer (DPB) to create a prediction of the currently-coded frame. Since the two input frames from DPB are available in both the encoder and the decoder, there is no need for side information to accomplish this task, so long as trained network operates at both the encoder and decoder. Fig.~\ref{fig:proposed_architecture} shows the overall architecture of the proposed network using two patches of the previous coded frames, $\widetilde{\mathbf{P}}_{t_{i}}$ as input, where $i \in \left\{1,2\right\}$. For bi-directional prediction, temporal relation among the inputs and the output is $t_1 < t < t_2$, where $\left|t-t_1\right| = \left|t-t_2\right| \leq 2$, meaning that the input frames are at the maximum temporal distance of $\pm2$ from the predicted frame. Meanwhile, for uni-directional prediction, $t_2=  t_1-1 = t-2$. 

Our proposed network inherits the U-Net structure to extract multi-scale deep features from the inputs, and produces spatially-varying filters through the \emph{Kernel Estimation} module in Fig.~\ref{fig:proposed_architecture}. In addition, inspired by~\cite{NIPSjaderberg2015spatial},  the \emph{Affine Estimation} module is introduced into the model to estimate affine motion parameters. Due to this motion information, the computed spatially-varying filters can be applied at  adaptively-chosen positions in the input patches. Moreover, excessively large filters are no longer needed to model the motion in the input patches. Indeed, the length of 1-D ASF in our model is 8 taps, which is the same as the length of the interpolation filter for motion estimation/compensation in  HEVC~\cite{hevc_mathias}. 
The U-Nets in both the Kernel Estimation and Affine Estimation module are made lighter (i.e., with fewer parameters) by using dilated convolutions~\cite{Yu2016MultiScaleCA}. Using the computed spatially-varying filters and affine motion information, two motion compensation operations -- motion compensation with local convolution (MCLC) and bilinear sampling -- generate four warped patches: $\mathbf{V}_{t_{i}}$ and $\mathbf{B}_{t_{i}}$, $i \in \left\{1,2\right\}$. Finally, the synthesis network (GridNet from~\cite{niklaus2018context}) merges the four warped patches to create an output patch, $\widehat{\mathbf{P}}_t$. 

\subsection{Light U-Net with dilated convolutions}
\label{ssec:u_net}

\begin{table}[t]
\centering
\caption{Depth-wise architectural details of the light U-Net}
\label{tbl:u_net_structure}
\smallskip\noindent
\resizebox{1\linewidth}{!}{%
\setlength\tabcolsep{3pt}
\renewcommand{\arraystretch}{1.5}
\begin{tabular}{@{}c|c|ccccc@{}}
\toprule
Depth (Side)                               & Input Dimension & \multicolumn{5}{c}{Channels or (Channels, Dilation)}        \\ \midrule
\multicolumn{1}{c|}{Depth1 (L)}       & $W\times H \times 3$        & 32  & 32  & (64,2)  & (96,4)  & 32  \\
\multicolumn{1}{c|}{Depth2 (L)}       & $\frac{W}{2} \times \frac{H}{2} \times 32$         & 64  & 64  & (96,2) & (128,4) & 64  \\
\multicolumn{1}{c|}{Depth3 (L)}       & $\frac{W}{4} \times \frac{H}{4} \times 64$         & 96 & 96 & (128,2) & (160,4) & 96 \\
\multicolumn{1}{c|}{Depth3 (R)}     & $\frac{W}{8} \times \frac{H}{8} \times 96$        & 96 & 96 & (128,2) & (160,4) & 96 \\
\multicolumn{1}{c|}{Depth2 (R)}       & $\frac{W}{4} \times \frac{H}{4} \times 96$        & 64  & 64  & (96,2) & (128,4) & 64  \\
\multicolumn{1}{c|}{Depth1 (R)}       & $\frac{W}{2} \times \frac{H}{2} \times 64$         & 32  & 32  & (64,2)  & (96,4) & 32  \\ \midrule
\multicolumn{2}{c|}{Output Feature Tensor}  & \multicolumn{5}{c}{$W \times H \times 64$}                  \\ \bottomrule
\end{tabular}}
\end{table}

Due to its multi-scale nature, a typical U-Net~\cite{Ronneberger2015UNetCN} architecture is quite demanding in terms of computation and memory. Since we use two U-Nets in our model, we need a more efficient solution. 
With reference to~\cite{Yu2016MultiScaleCA}, we rebuild the U-Net from~\cite{deep_frame_prediction} using dilated convolutions, and refer to it as \emph{light U-Net} here. Table~\ref{tbl:u_net_structure} shows details of the light U-Net. The second column shows input dimensions (width $\times$ height $\times$ channels) at each depth. There are five convolutional layers at each depth, each with $3\times3$ kernels and \texttt{LeakyReLU} activation; the first, second, and fifth are regular convolutions, while the third and fourth are dilated convolutions. The five entries in the third column in the table specify the number of channels at the output of each convolutional layer, and in the case of dilated convolutions, (channels, stride) are listed. 
The first three rows show the left side of the light U-Net, where average 2$\times$2 pooling decreases the width and the height by a factor of 2 at each depth. The next three rows show the right side of the network, where  upsampling is performed via bilinear interpolation prior to convolution. 

Accounting for dilation and pooling, the receptive field at Depth1 (L) is $19\times19$, but at Depth3 (R) it reaches $152\times152$, the largest value in the model. 
Therefore, for the size of input patches, we use $W=H=152$ during both training and testing. This can be considered as being analogous to the maximum search range in conventional motion estimation. 

\subsection{Kernel estimation}
\label{ssec:kernel_estimation}

Our kernel estimation module is built upon the light U-Net and produces adaptive 1-D separable filters (ASFs). Using two input patches, $\widetilde{\mathbf{P}}_{t_i}$, 
a feature tensor with size of $W \times H \times 64$ is produced at the output of the U-Net, then fed into ASF blocks as shown in Fig.~\ref{fig:proposed_architecture}. The ASF blocks are fully-convolutional, with four layers using $3\times3$ filters. The first three convolutional layers use \texttt{LeakyReLU} activation, while the last one uses linear activation. 
Four output tensors are produced: $\mathbf{F}_{t_i}^{d} \in \mathbb{R}^{W \times H \times C}$, where $d \in \{h, v\}$, to create spatially-varying ASFs. Each of these tensors contains 1-D filter coefficients along the channel axis at location $(x,y)$, and $C$ represents the filter length. In order to create the spatially-varying 2-D convolution filter kernel for the pixel location $(x,y)$, horizontal ($h$) and vertical ($v$) filters are multiplied using outer product: 
\begin{equation}
    \mathbf{K}_{t_i}^{(x,y)} = \mathbf{F}_{t_i}^{h}(x,y,:) \otimes \mathbf{F}_{t_i}^{v}(x,y,:) = \mathbf{F}_{t_i}^{v}(x,y,:)\mathbf{F}_{t_i}^{h}(x,y,:)^T
\label{eq:kernel_k}
\end{equation}
\noindent Filters $\mathbf{K}_{t_i}^{(x,y)}$ are subsequently applied to input patches to produce warped patches. Regarding the filter length, we tried  using different $C \in \{8, 9, 13, 17, 21, 25\}$, but none showed noticeable difference in performance compared to others, so we settled on $C=8$. Due to the motion-adaptive selection of the location where to apply filters, to be explained below, longer filters don't seem to offer further gain. It is interesting to note that HEVC also uses 8-tap DCT-based  interpolation filters in motion compensation~\cite{hevc_mathias}.

\subsection{Affine parameter estimation}
\label{ssec:affine_estimation}

Architecturally, the affine estimation module resembles the kernel estimation module, and is built upon the light U-Net. The feature tensor produced from two input patches 
is fed into two affine parameter estimation blocks, whose structure is the same as ASF blocks, except for \texttt{Tanh} activation instead of \texttt{LeakyReLU}. These blocks generate affine transformation parameters, $\mathbf{\Theta}_{t_i} \in \mathbb{R}^{W \times H \times N}$ as shown in Fig.~\ref{fig:proposed_architecture}. Note that the model estimates one set of motion parameters per pixel, rather than a block of pixels. We experimented with various affine models, and found that with more than three parameters, the training becomes unstable, likely due to overfitting. For this reason we also excluded rotation from the motion model.  
Hence, only translation and isotropic scaling parameters are considered in the affine transformation, so $N=3$ parameters are produced for each pixel location $(x,y)$: $\theta_{t_i,1}$ for isotropic scaling, $\theta_{t_i,2}$ for horizontal translation, and $\theta_{t_i,3}$ for vertical translation. 
Using these parameters, a $2\times3$ affine transformation matrix can be formed for each $(x,y)$ and each $t_i$: 
\begin{equation}
    \mathbf{A}_{t_i}(x,y) =
    \begin{bmatrix}
\theta_{t_i, 1} & 0 & \theta_{t_i, 2}\\ 
0 & \theta_{t_i, 1} & \theta_{t_i, 3}
\end{bmatrix}.
\label{eq:affine_parameters_vector}
\end{equation}

Inspired by~\cite{NIPSjaderberg2015spatial}, in order to compute the spatial affine coordinate transformation, we introduce the grid generator function $\mathcal{T}_{\mathbf{G}}$ with respect to $\mathbf{\Theta}_{t_{i}}$ at location $(x,y)$: 
\begin{equation}
\mathbf{S}_{t_i}(x,y,:)=\mathcal{T}_{\mathbf{G}}(\mathbf{\Theta}_{t_i};x,y)  = 
\mathbf{A}_{t_i}(x,y)
\begin{bmatrix}
\mathbf{G}(x,y,0)\\ 
\mathbf{G}(x,y,1)\\ 
1
\end{bmatrix},
\label{eq:final_grid}
\end{equation}
\noindent where $\mathbf{G}(:,:,0)$ and $\mathbf{G}(:,:,1)$ represent the normalized horizontal and vertical coordinates, respectively, such that $-1 \leq \mathbf{G}(x, y, :) \leq 1$. The final $\mathbf{S}_{t_i} \in \mathbb{R}^{W \times H \times 2}$ shows how location $(x,y)$ is warped according to the estimated affine parameters: $\mathbf{S}_{t_i}(x,y,1)$ is the new horizontal coordinate, and $\mathbf{S}_{t_i}(x,y,2)$ is the new vertical coordinate. 

\subsection{Motion compensation: MCLC and Bilinear sampling}
\label{ssec:motion_compensations}

With the computed spatially-varying 2-D filter kernels $\mathbf{K}_{t_i}$ and the spatial transformation coordinates $\mathbf{S}_{t_i}$, two motion compensation operations -- motion compensation with local convolution (MCLC) and bilinear sampling -- are conducted to create four warped patches: $\mathbf{V}_{t_i}$ and $\mathbf{B}_{t_i}$, for $i\in\{1,2\}$. First, MCLC predicts a pixel value at $(x,y)$ for each color channel $c \in \{1, 2, 3\}$, by applying filter $\mathbf{K}_{t_i}^{(x,y)}$ to the area centered around $\mathbf{S}_{t_i}(x,y,:)$ in the input patch $\widetilde{\mathbf{P}}_{t_i}$. 
Specifically,
\begin{equation}
\mathbf{V}_{t_i}(x,y,c) = \sum \mathbf{K}^{(x,y)}_{t_1} \circ \widetilde{\mathbf{P}}^{\left(\left\lfloor n_{\mathbf{S}_{t_i}}^{(x,y)} \right\rfloor, \left\lfloor m_{\mathbf{S}_{t_i}}^{(x,y)}\right\rfloor\right)}_{t_i}(:,:,c),
\label{eq:mclc_patch}
\end{equation}
where $\lfloor\cdot\rfloor$ represents flooring to the nearest integer, while $n_{\mathbf{S}_{t_i}}^{(x,y)}$ and $m_{\mathbf{S}_{t_i}}^{(x,y)}$ are 
un-normalized coordinates corresponding to $\mathbf{S}_{t_i}(x,y,:)$. $\widetilde{\mathbf{P}}^{(\lfloor n_{\mathbf{S}_{t_i}}^{(x,y)} \rfloor, \lfloor m_{\mathbf{S}_{t_i}}^{(x,y)}\rfloor)}_{t_i}(:,:,c)$ is a $C \times C$ region of the corresponding input patch centered at $(\lfloor n_{\mathbf{S}_{t_i}}^{(x,y)} \rfloor, \lfloor m_{\mathbf{S}_{t_i}}^{(x,y)}\rfloor)$ in color channel $c$, and  $\circ$ denotes the Hadamard product (element-wise multiplication). The sum ($\sum$) goes over all elements in the Hadamard product. Near the patch boundary where there may be insufficient input samples, the $C \times C$ matrix is filled up with the nearest-neighboring samples. 

In~(\ref{eq:mclc_patch}), filter kernel $\mathbf{K}^{(x,y)}_{t_1}$ is produced using trainable parameters, and the loss gradient with respect to these parameters is computed  following~\cite{Niklaus_ICCV_2017}. However, the location where the filter is applied 
is also computed using trainable parameters involved in producing $\mathbf{S}_{t_i}$, and computing the loss gradient relative to these parameters requires differentiating  $\mathbf{V}_{t_i}$ with respect to $x$ and $y$. Since these are discrete quantities, we use 
the discrete approximation to the partial derivative:
\begin{equation}
\frac{\partial\mathbf{V}_{t_i}}{\partial x} \approx \mathbf{V}_{t_i}(x+1, y, c) - \mathbf{V}_{t_i}(x, y, c).
\label{eq:mclc_patch_backward}
\end{equation}
\noindent The derivative with respect to $y$ is obtained similarly. Using these approximations, standard backpropagation can
be used for training the model. 

In addition to $\mathbf{V}_{t_i}$, we produce another pair of warped patches, $\mathbf{B}_{t_i}$, from affine parameters, using bilinear interpolation. This may be considered as a ``skip'' connection in motion compensation, using only the necessary grid mapping via bilinear interpolation, but skipping the filtering operation performed in~(\ref{eq:mclc_patch}). 
Specifically,
\begin{equation}
\begin{split}
\mathbf{B}_{t_i}(x,y,c) = \sum_{p=0}^{1}\sum_{q=0}^{1} 
&\max\left\{0, 1-(\lfloor n_{\mathbf{S}_{t_i}}^{(x,y)} \rfloor + p + n_{\mathbf{S}_{t_i}}^{(x,y)})\right\} \cdot\ \\[1ex] 
&\max\left\{0, 1-(\lfloor m_{\mathbf{S}_{t_i}}^{(x,y)} \rfloor + q + m_{\mathbf{S}_{t_i}}^{(x,y)})\right\} \cdot \ \\[1ex]
&\widetilde{\mathbf{P}}_{t_i}\left(\lfloor n_{\mathbf{S}_{t_i}}^{(x,y)} \rfloor + p, \lfloor m_{\mathbf{S}_{t_i}}^{(x,y)} \rfloor + q, c \right)
\end{split}
\label{eq:bilinear_sampling}
\end{equation}
\noindent where $p$ and $q$ are indices to access four neighbouring pixels and used to compute the proportion of distance between pixel locations, and the output pixel value is sum of weighted neighboring pixels. Since~(\ref{eq:bilinear_sampling}) is sub-differentiable~\cite{NIPSjaderberg2015spatial, bao2019memc}, the loss gradient can flow back to the affine transformation module during backpropagation. Overall, 
four warped patches, $\mathbf{V}_{t_i}$ and $\mathbf{B}_{t_i}$, are constructed via motion compensation and passed on to the synthesis module (Fig.~\ref{fig:proposed_architecture}). 

\subsection{Synthesis}
\label{ssec:synthesis}
\begin{figure}[t]
    \centering
    \begin{minipage}[b]{1\linewidth}
    \centering
    \includegraphics[width=\textwidth]{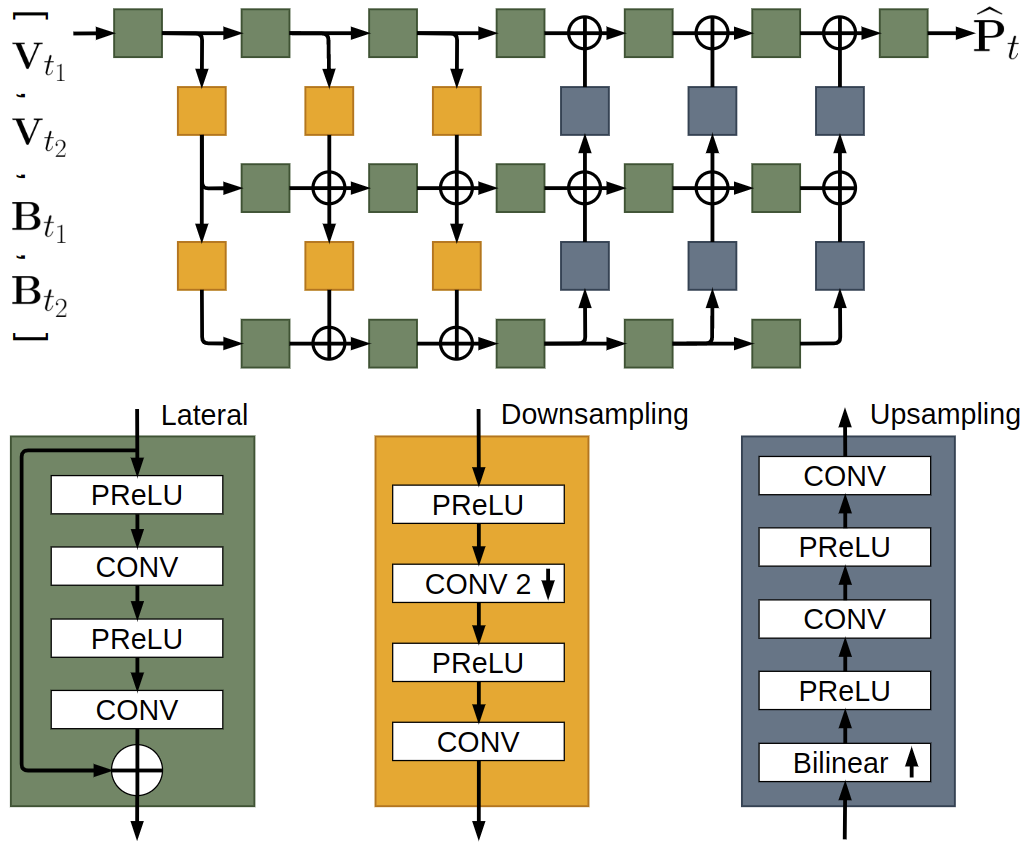}
    \end{minipage}
\caption{Architecture of a modified GridNet~\cite{niklaus2018context}, combining four pre-warped patches (stacked together at input) to create an output patch $\widehat{\mathbf{P}}_{t}$ 
}
\label{fig:synthesis_architecture}
\end{figure}


The four motion-compensated patches, $\mathbf{V}_{t_i}$ and $\mathbf{B}_{t_i}$, $i \in \{1, 2\}$, are stacked together and used to produce the output patch $\widehat{\mathbf{P}}_t$ using GridNet~\cite{niklaus2018context}, as shown in Fig.~\ref{fig:synthesis_architecture}.
GridNet is composed of three rows of processing blocks, referred to as feature streams, and processes the input patches with constant resolution in each row. Features from various streams are exchanged at different scales through down- and up-sampling blocks, with the idea that local high-resolution predictions are guided by global low-resolution information~\cite{niklaus2018context, xia2019deep}. The details of each processing block, borrowed from~\cite{niklaus2018context}, are depicted at the bottom of Fig.~\ref{fig:synthesis_architecture}.
Feature resolution is reduced by half, horizontally and vertically using a convolutional layer with a stride of 2, when passed on to the lower resolution stream, and conversely, increased by a factor of two using bilinear up-sampling, when passed on to the higher resolution stream.   
Overall, GridNet provides a powerful and flexible architecture for trainable merging of input patches into the final patch $\widehat{\mathbf{P}}_t$.

\subsection{Loss function}
\label{ssec:loss_function}

Considering the fact that output patches are to be used as predictors in a video compression pipeline, 
when building the loss function for training the model, we focus on coding-relevant metrics, rather than perceptual metrics. 
In practice, Sum of Absolute Transformed Difference (SATD) is used in conventional video coding, with Hadamard transform as a proxy for DCT due to its lower computational cost. However, since we train our DNN offline, high computational cost is acceptable at training time. Therefore, we use  DCT-based SATD loss terms during training. Considering various transform sizes used in practical video coding, we also measure SATD with different sizes of DCT ($8\times8$, $16\times16$, and $32\times32$) on 
the residual signal $\mathbf{R}_t = \widehat{\mathbf{P}}_{t} - \mathbf{P}_{t}$ for each transform size. 
Let $\mathbf{R}_{t,k}^{j}$ be 
the residual block with size  $j\times j$, where $j\in\{8, 16, 32\}$, and $k$ represents the block index, $k \in \{1, 2, 3, ..., \frac{W\cdot H}{j^2}\}$. Then, the SATD loss term is formed by
\begin{equation}
\mathcal{L}_{\textup{SATD}, j}= \sum_{k=1}^{W\cdot H/j^2}\norm{\mathbf{D}^j \mathbf{R}^{j}_{t,k}(\mathbf{D}^{j})^T}_{1}, 
\label{eq:l1_satd_loss}
\end{equation}
\noindent where $\mathbf{D}^{j}$ represents the DCT matrix for $j\times j$ DCT. 

The second loss term is based on the scaled Mean Squared Error (MSE) between the generated output patch and ground truth at multiple scales:
\begin{equation}
\mathcal{L}_{\textup{MSE}, s}= \norm{\widehat{\mathbf{P}}_{t}^{1/s} - \mathbf{P}_{t}^{1/s}}_{2}^{2},
\label{eq:MSE_loss}
\end{equation}
\noindent where $s \in \{1, 2, 4\}$,  $\widehat{\mathbf{P}}_t^{1/s}$ represents the generated patch at scale $1/s$ (both vertically and horizontally), while $\mathbf{P}_t^{1/s}$ is the ground truth patch at the same scale. Bilinear interpolation is used to perform rescaling to the desired resolution. 

Finally, we use an object context loss term $\mathcal{L}_{\textup{CTX}}$ by measuring feature reconstruction error in an object detection backbone. Feature losses based on image classification backbones (e.g., VGG) are quite popular. However, since image classification is invariant to small shifts, it can be expected that for successful prediction in video coding, an object detection backbone may give a more useful set of features, because object detection involves bith classification and spatial localization. 
We define the context loss term as
\begin{equation}
\mathcal{L}_{\textup{CTX}}=\norm{\phi(\widehat{\mathbf{P}}_{t}) - \phi(\mathbf{P}_{t})}_{2}^{2},
\label{eq:feature_loss}
\end{equation}
\noindent where $\phi(\cdot)$ is an object context function that extracts features from the 12-th layer in YOLO version 3~\cite{redmon2018yolov3}, using $\mathbf{P}_{t}$ and $\widehat{\mathbf{P}}_{t}$ as inputs to the detector.


Finally, the overall loss function is a combination of the three types of loss terms~(\ref{eq:l1_satd_loss})--(\ref{eq:feature_loss}):
\begin{equation}
\mathcal{L} =\sum_{j \in \{8, 16, 32\}}\mathcal{L}_{\textup{SATD}, j} + \sum_{s \in \{1, 2, 4\}} s^2 \cdot \mathcal{L}_{\textup{MSE}, s} + \mathcal{L}_{\textup{CTX}},  
\label{eq:final_loss}
\end{equation}
\noindent where each scale of the MSE loss is weighted proportionally to the scale $s$.

\subsection{Training and ablation study}
\label{ssec:training_strategy}

Our frame prediction network is implemented in PyTorch and trained with the AdaMax~\cite{kingma2014adam} optimizer by following the strategy described in~\cite{deep_frame_prediction}. Additionally, we further train the network on the VIMEO-90K dataset~\cite{xue2019video}. All training triplets (two input patches and one output patch) are prepared in the YUV420 format before training. During training, randomly selected triplets 
with size of 152$\times$152 are presented to the network in batches of 16. We train models separately for bi- and uni-directional prediction. Depending on the prediction direction, triplets are properly assigned to input and output with data augmentation schemes~\cite{Niklaus_ICCV_2017}. 

\begin{figure}[t]
    \centering
    \begin{minipage}[b]{1\linewidth}
    \centering
    \includegraphics[width=\textwidth]{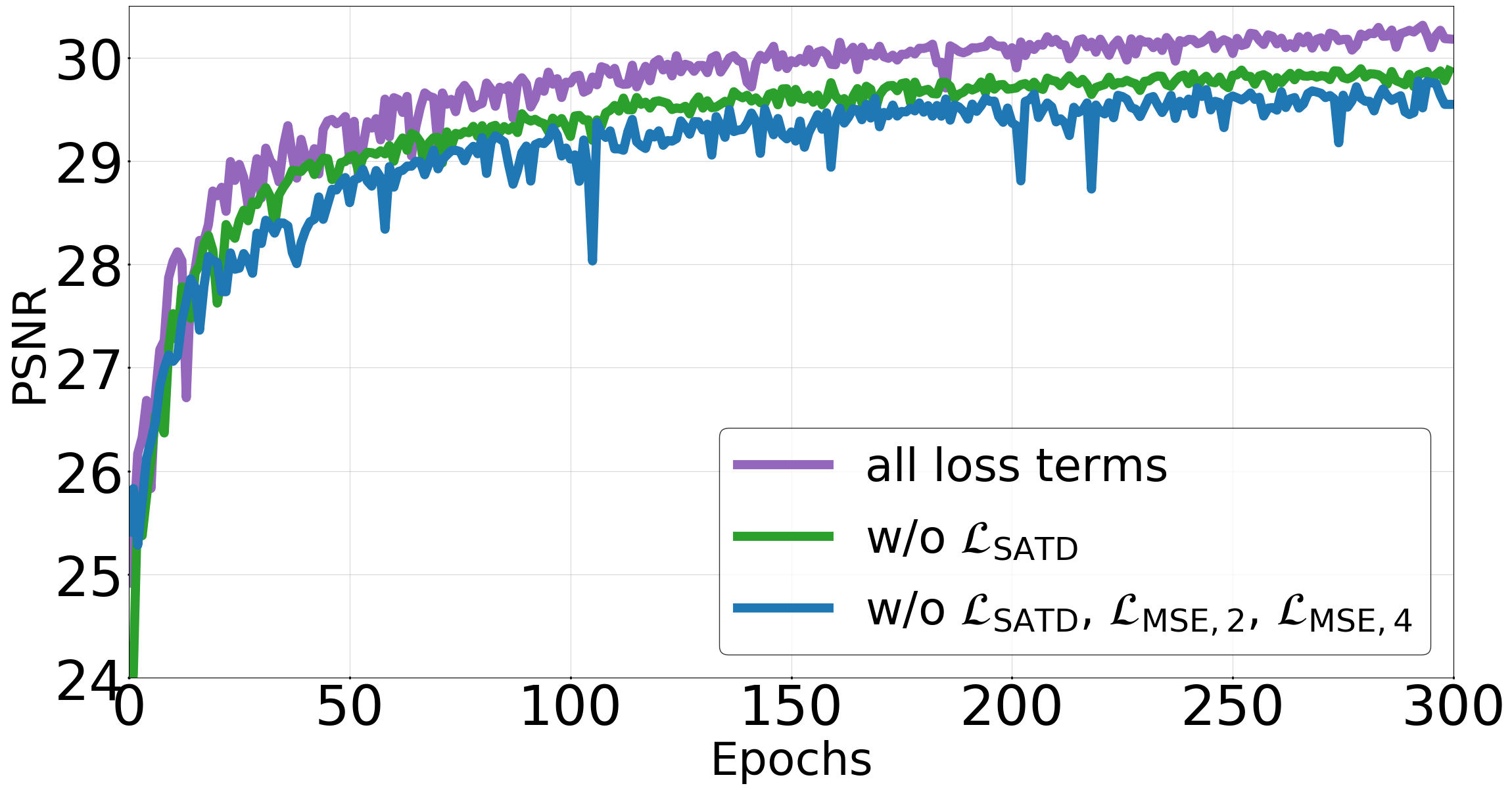}
    \end{minipage}
\caption{Evolution of validation Y-PSNR of our bi-prediction network for various versions of the ablated loss function.}
\label{fig:ablation_study}
\end{figure}

In the first training stage, using randomly selected 6,686 triplets from 24 sequences with CIF resolution as input, our network is trained for 300 epochs, with an initial learning rate of 0.001 decreased by polynomial decay every 5 epochs. 
In this stage we also conduct an ablation study to examine the efficacy of various loss terms. In particular, we examine the performance of the network with $\mathcal{L}_{\textup{SATD},j}$ for all $j$  
and $\mathcal{L}_{\textup{MSE}, s}$ for $s\in\{2,4\}$ removed. Fig.~\ref{fig:ablation_study} shows validation Y-PSNR of our bi-prediction network depending for various loss functions. The purple curve represents the case of using all loss terms in~(\ref{eq:final_loss}). The green curve is the loss function~(\ref{eq:final_loss}), but with $\mathcal{L}_{\textup{SATD},j}$ removed, while the blue curve corresponds to further removal of $\mathcal{L}_{\textup{MSE}, s}, s\in\{2,4\}$ from the loss function. As seen in the figure, the terms $\mathcal{L}_{\textup{MSE}, s}, s\in\{2,4\}$ provide an improvement of about 0.2 dB in PSNR, while $\mathcal{L}_{\textup{SATD},j}$ terms bring an additional 0.5 dB improvement. 
Therefore, we keep all lost terms in~(\ref{eq:final_loss}) in subsequent training.

In the second stage, we continue training from the network weights obtained in the previous stage, with a learning rate of 0.0005 on the dataset including 27,360 triplets generated by~\cite{deep_frame_prediction}. While training our model for 600 epochs, all the augmentation strategies  from~\cite{Niklaus_ICCV_2017} are used. We also use those augmentation methods in the last (third) training stage, which is carried out on the VIMEO-90K dataset, with a learning rate of 0.0001 gradually reduced for 300 epochs. In total, 55,095 triplet patches are randomly selected from the dataset during training. The model was trained on a GeForce RTX 2080 GPU with 11 GiB RAM and it took about a week to complete all three training stages.

\section{Integration into HEVC}
\label{sec:network_strategies}

\subsection{Patch-based frame composition}

\begin{figure}[t]
    \centering
    \begin{minipage}[b]{1\linewidth}
    \centering
    \includegraphics[width=\textwidth]{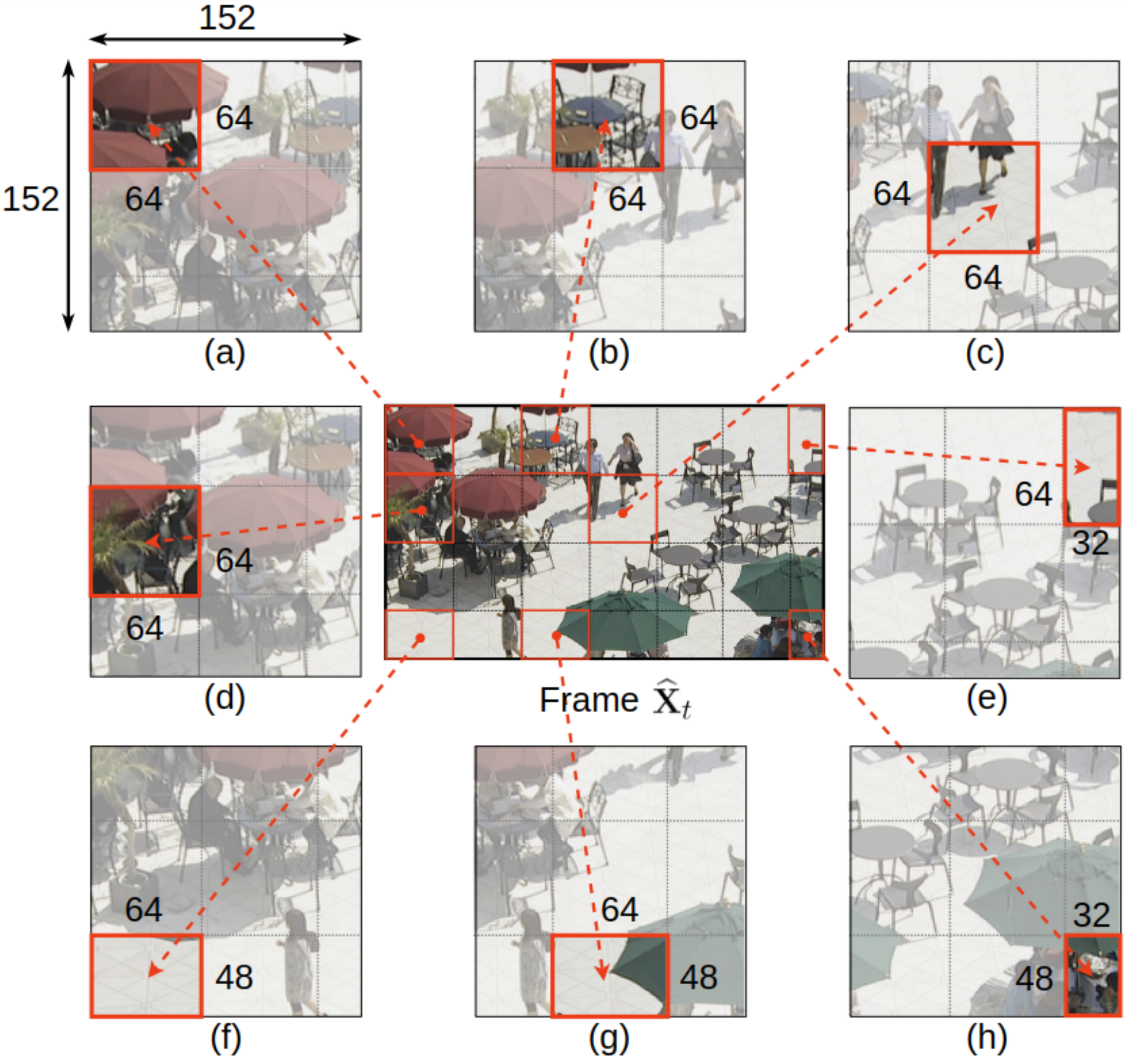}
    \end{minipage}
\caption{An example of patch composition to create an entire predicted frame with the size of 416$\times$240}
\label{fig:patchwork_example}
\end{figure}

In this section, we explain how we form the predicted frame from the predicted patches. As discussed in Section~\ref{ssec:u_net}, patches of size 152$\times$152 are used as inputs to the DNN, to produce an output patch $\widehat{\mathbf{P}}_t$ of the same size. Patches are selected from input frames using a sliding-widow approach with a stride of 64, except near frame boundaries. The first pair of patches is selected from the top-left corner of the input frames. The next pair of patches is selected with a stride of 20, and subsequent ones with the stride of 64. Once the bottom and right frame boundaries are reached, the patches are selected so that the patch boundary matches the frame boundary, since no padding is used.    

To create the entire predicted frame $\widehat{\mathbf{X}}_t$, we combine the output patches as shown in Fig.~\ref{fig:patchwork_example}. 
From each output patch $\widehat{\mathbf{P}}_t$ we take an area of size at most 
64$\times$64 
to place it in the corresponding location in $\widehat{\mathbf{X}}_t$, starting with top left. The procedure is shown in Fig.~\ref{fig:patchwork_example} for blocks around the frame boundary. The top-left 64$\times$64 block comes from the first predicted patch (Fig.~\ref{fig:patchwork_example}(a)), the next 64$\times$64 block comes from the second predicted patch (Fig.~\ref{fig:patchwork_example}(b)), and so on. Near the right and bottom frame boundary, an area of appropriate size is selected from the corresponding predicted patch, as shown in Fig.~\ref{fig:patchwork_example}(e)-(h). 
For the inner blocks, we use the area at the center of the corresponding output patch, as shown in Fig.~\ref{fig:patchwork_example}(c).

\subsection{Proposed integration method}
\label{sec:integration}

\begin{figure}[t]
    \centering
    \begin{minipage}[b]{1\linewidth}
    \centering
    \includegraphics[width=\textwidth]{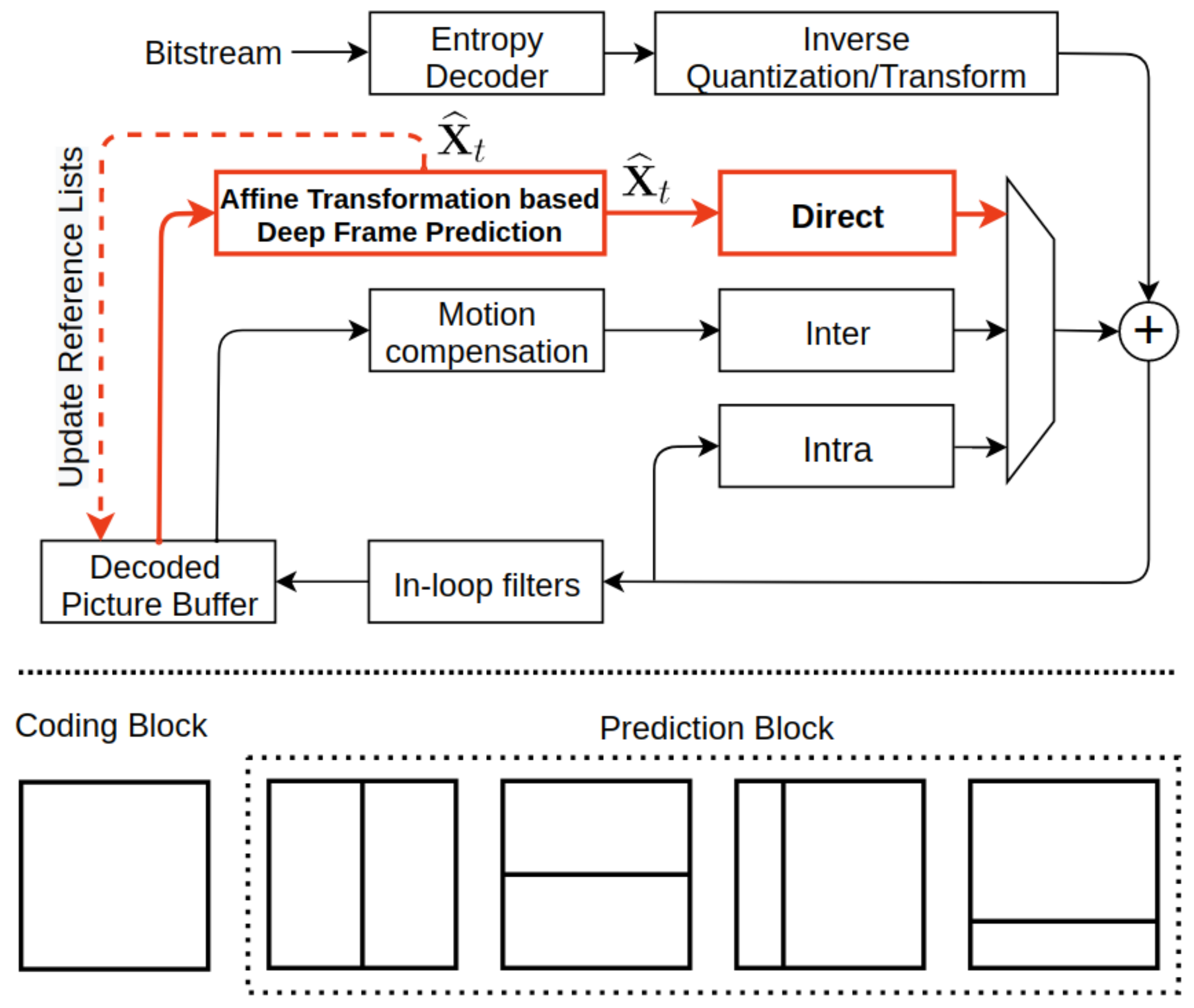}
    \end{minipage}
\caption{Integrated proposed method in the decoder and shapes of coding and prediction blocks in HEVC}
\label{fig:coding_and_prediction_blocks}
\end{figure}
Several methods have been proposed to integrate predicted frames into an HEVC coding pipeline~\cite{deep_frame_prediction, xia2019deep, zhao2019enhanced, hevc_with_sep_conv_for_ra}. Here we propose a method that effectively combines two different approaches in a rate-distortion optimized (RDO) manner, such that the benefits of both could contribute to coding efficiency. Specifically, one method~\cite{deep_frame_prediction} is to use DNN-generated frames with an additional prediction mode, referred to as ``Direct'' mode in Fig.~\ref{fig:coding_and_prediction_blocks}. The direct mode competes with all other coding modes (i.e., intra and inter prediction) in the RDO mode selection in the encoder. When the direct mode is used, a flag is set so that the decoder knows to compensate the corresponding square block from the collocated block in the DNN-generated frame $\widehat{\mathbf{X}}_t$. 
Note that this flag is only valid when the direct mode is feasible, i.e., when $\widehat{\mathbf{X}}_t$ exists, to minimize overhead. For example, 
when there are no available reference frames to create the DNN-predicted frame $\widehat{\mathbf{X}}_t$, 
the direct mode flag is unavailable. After compensating a block from the DNN-generated frame, residuals are added to 
to reconstruct the pixels. 
The DNN-predicted frame already includes motion warping, so no additional motion information needs to be transferred -- this is where most of the coding gain of this approach comes from~\cite{deep_frame_prediction}. 
However, direct mode also restricts DNN-frame information to be used only in square blocks; enabling additional block shapes would require additional flag bits. 

\begin{figure}[t]
    \centering
    \begin{minipage}[b]{0.95\linewidth}
    \centering
    \includegraphics[width=\textwidth]{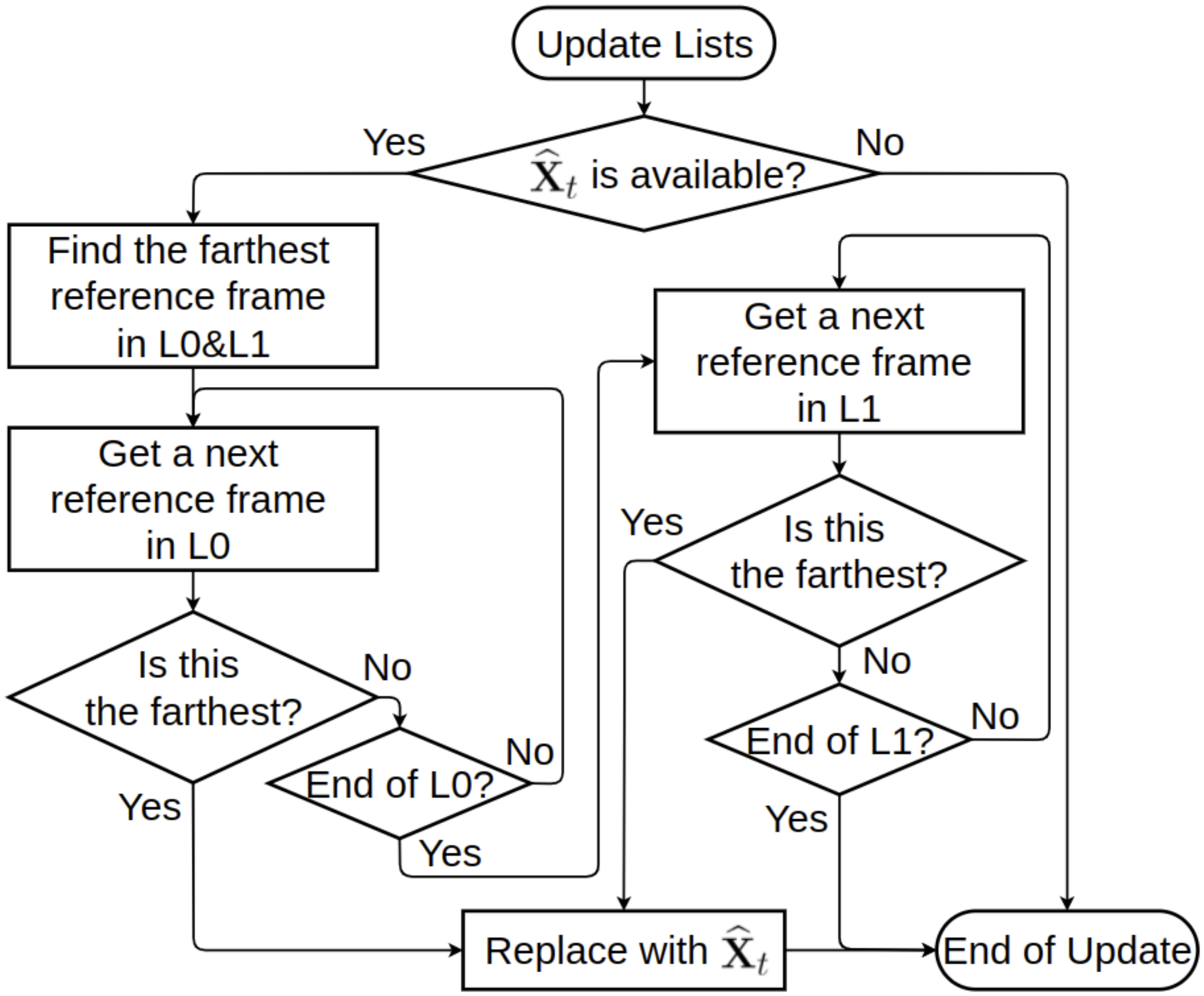}
    \end{minipage}
\caption{Updating reference lists for bi-directional prediction 
}
\label{fig:update_ref_list}
\end{figure}

To overcome the above mentioned restriction of the direct mode, we supplement it with another integration method from~\cite{xia2019deep}, which is 
indicated by a dashed red line in Fig.~\ref{fig:coding_and_prediction_blocks}. In this approach, 
the current reference list is updated by replacing one of the frames with the DNN-generated frame $\widehat{\mathbf{X}}_t$, and then treating this frame as any other reference frame for inter prediction. In this approach, motion information may still need to be encoded if $\widehat{\mathbf{X}}_t$ is used as a reference, but it is now possible to use a wider variety of prediction block shapes, as indicated in the bottom part of Fig.~\ref{fig:coding_and_prediction_blocks}. 

Regarding the reference list update, 
two cases need to be considered. For bi-directional prediction, two reference lists -- \texttt{L0}, comprising decoded frames temporally preceding the current frame, 
and \texttt{L1}, comprising decoded frames temporally succeeding the current frame -- 
are used. In certain special cases, the same frame could appear in both lists.  Fig.~\ref{fig:update_ref_list} presents the process of updating reference lists for the bi-directional prediction. This flow is only viable in the case that there are proper reference frames at the maximum temporal distance of $\pm2$ from the current frame, such that our DNN is able to create $\widehat{\mathbf{X}}_t$. Otherwise, inter prediction makes use of the default reference lists without update. When the DNN-generated frame is available, we find the decoded frame with the largest POC difference from the current frame among both lists. Three cases can arise: (1) this frame is in \texttt{L0}; (2) this frame is in both \texttt{L0} and \texttt{L1}; (3) this frame is in \texttt{L1}. In the first two cases, the frame in \texttt{L0} is replaced by the DNN-generated frame $\widehat{\mathbf{X}}_t$, and in the third case, the frame in \texttt{L1} is replaced. 

For uni-directional prediction, temporally preceding decoded frames, starting with the frame nearest to the current frame, are placed in the reference list. Hence, updating the reference list for uni-directional prediction is much simpler. We always replace the third reference frame in the list with $\widehat{\mathbf{X}}_t$ when there are three or more frames in the lists. The rationale is that  the first two reference frames are needed as inputs to the DNN to create $\widehat{\mathbf{X}}_t$, because they are temporally closest frames to the current frame, so they are left in place. On the other hand, choosing a frame farther than the third to replace by $\widehat{\mathbf{X}}_t$ would end up costing more bits to indicate when $\widehat{\mathbf{X}}_t$ is used as a reference. 
Indeed, based in the results in Section~\ref{ssec:coding_performance_of_two}, we find that the DNN-generated frame ends up being used as a reference up to 14 times more frequently than the default third frame, depending on the sequence. Further details will be provided with the discussion of Fig.~\ref{fig:comp_r3_ref_analysis} in Section~\ref{ssec:coding_performance_of_two}.

Regarding the compatibility with existing motion-related tools in HEVC, such as \textup{MERGE} and \textup{AMVP}~\cite{hevc_mathias}, we follow the strategy used in our previous work~\cite{deep_frame_prediction} when a neighboring block is coded with the direct mode. Specifically, the blocks with direct mode are excluded from all motion-related candidate lists. On the other hand, when a neighboring block refers to a DNN-generated frame in reference lists using conventional inter prediction, that block is considered as a candidate in a similar manner to~\cite{lee2020deep}. One difference compared to~\cite{lee2020deep} is that in our method, the block does not participate in temporal motion vector search in \textup{AMVP}. Additionally, we also tried converting the motion information produced by our DNN into motion vector-like information to be used by these motion prediction tools, however, this did not improve the coding efficiency. A possible reason is that our DNN produces motion that attempts to minimize the loss function~(\ref{eq:final_loss}), whereas conventional MEMC is tailored to reducing the RD cost. Therefore, when there is a neighboring block using the DNN-generated information, the block is excluded from any of the motion-related prediction tools. Finally, during encoding, the new coding modes involving the DNN-generated frame enter the RDO process and compete with the conventional coding modes within HEVC, without any extra advantage.

\section{Experimental results}
\label{sec:experimental_results}
In this section, we first show visualizations of the estimated affine motion for uni- and bi-directional frame prediction. 
Then, we evaluate the coding performance of the proposed methods integrated into HEVC test model (HM)~\cite{hm} according to Section~\ref{sec:network_strategies}. To facilitate comparison with existing methods in the literature, we integrate the proposed methods into HM-16.6 and HM-16.20. Note that these two versions of HM use different encoding parameters associated with the reference list and coding structure, which 
affects the rate-distortion (RD) performance. We followed the common test conditions~\cite{hevc_ctc} to measure coding performance with four quantization parameters QP $\in \{22, 27, 32, 37\}$ for Low delay (LD), Low delay P (LDP), and Random access (RA) configurations for inter frame coding. 

\subsection{Visualizations of the learned motion field}

\begin{figure*}[t]
    \centering
    \begin{minipage}[b]{0.24\linewidth}
    \centering
    \includegraphics[width=\textwidth]{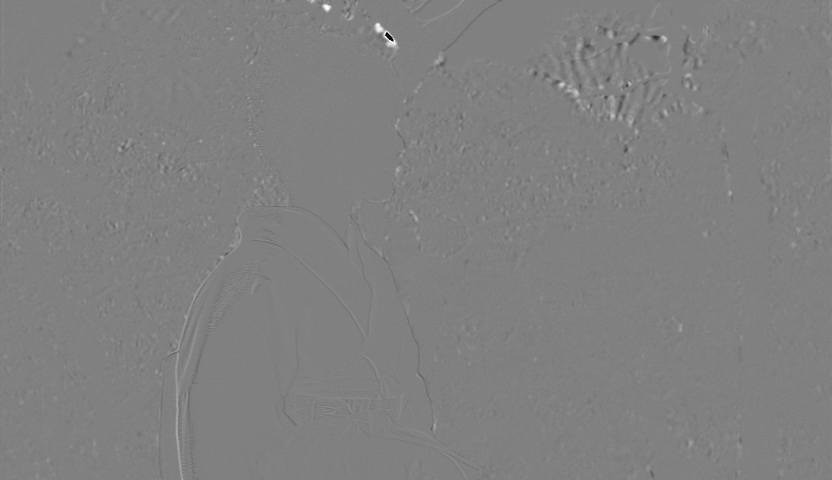}
    \end{minipage}
    \begin{minipage}[b]{0.24\linewidth}
    \centering
    \includegraphics[width=\textwidth]{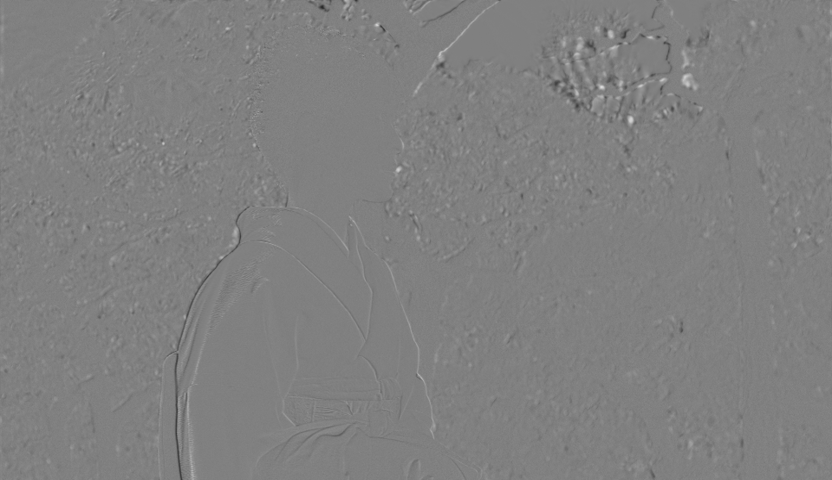}
    \end{minipage}
    \begin{minipage}[b]{0.24\linewidth}
    \centering
    \includegraphics[width=\textwidth]{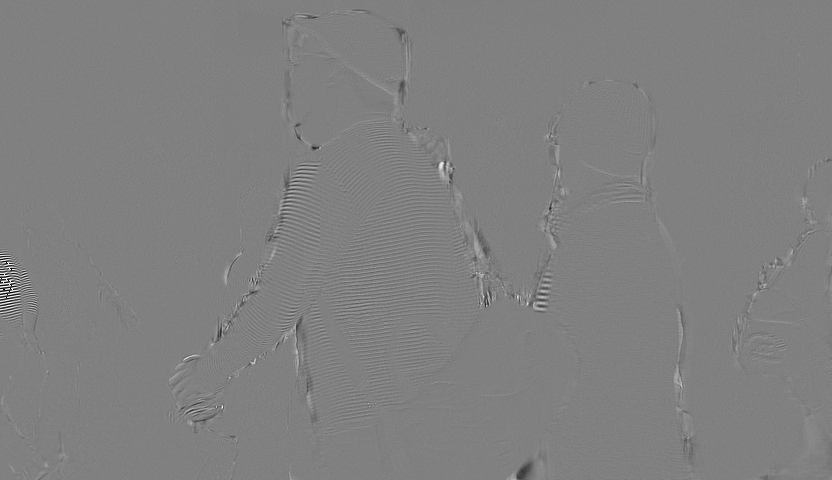}
    \end{minipage}
    \begin{minipage}[b]{0.24\linewidth}
    \centering
    \includegraphics[width=\textwidth]{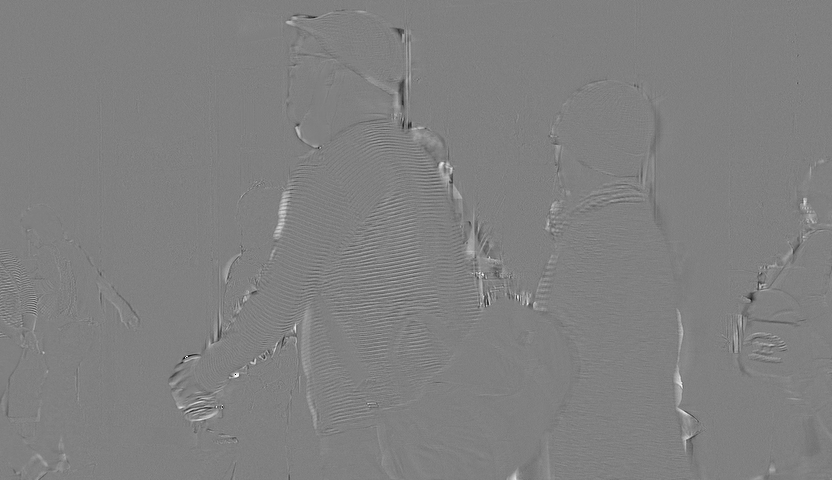}
    \end{minipage}
    \hfill
    \vspace{0.12cm}

    \begin{minipage}[b]{0.24\linewidth}
    \centering
    \includegraphics[width=\textwidth]{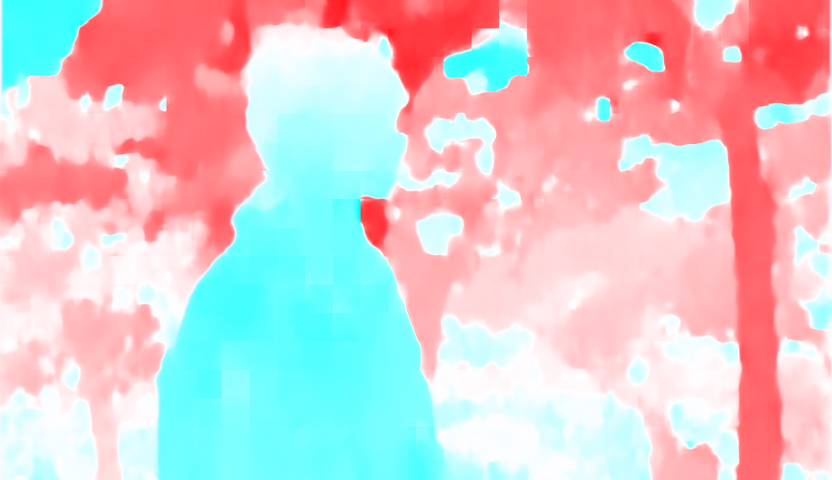}
    \end{minipage}
    \begin{minipage}[b]{0.24\linewidth}
    \centering
    \includegraphics[width=\textwidth]{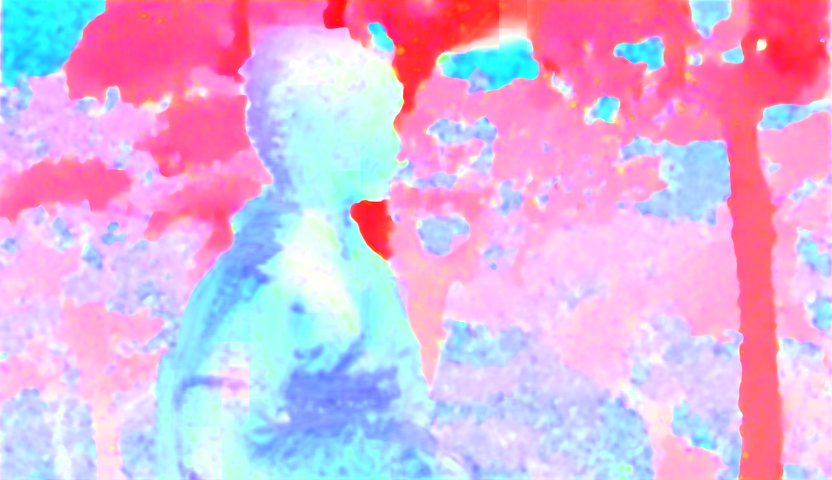}
    \end{minipage}
    \begin{minipage}[b]{0.24\linewidth}
    \centering
    \includegraphics[width=\textwidth]{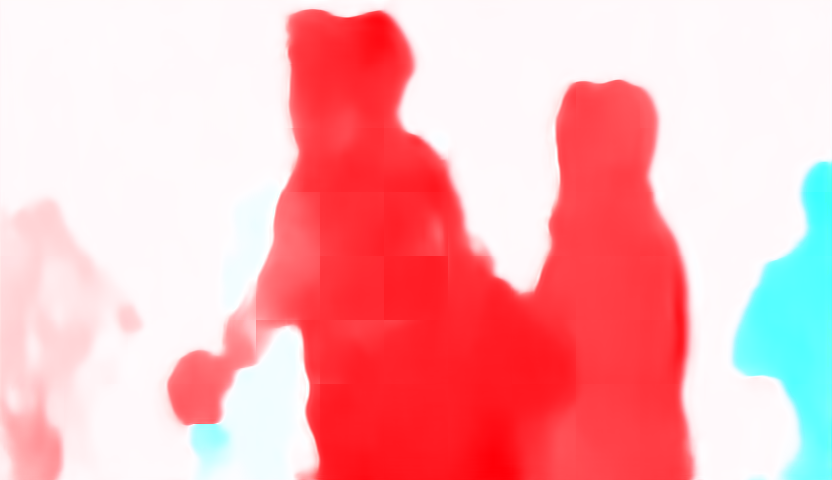}
    \end{minipage}
    \begin{minipage}[b]{0.24\linewidth}
    \centering
    \includegraphics[width=\textwidth]{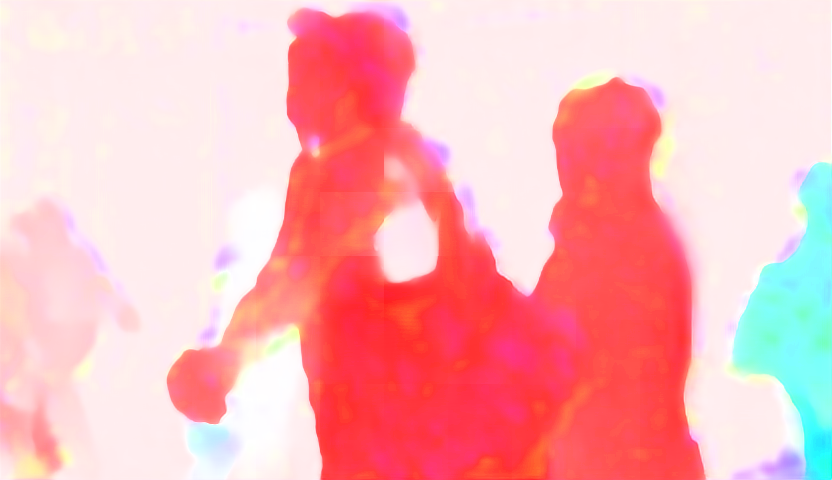}
    \end{minipage}
    \hfill
    \vspace{0.12cm}
    
    \begin{minipage}[b]{0.24\linewidth}
    \centering
    \includegraphics[width=\textwidth]{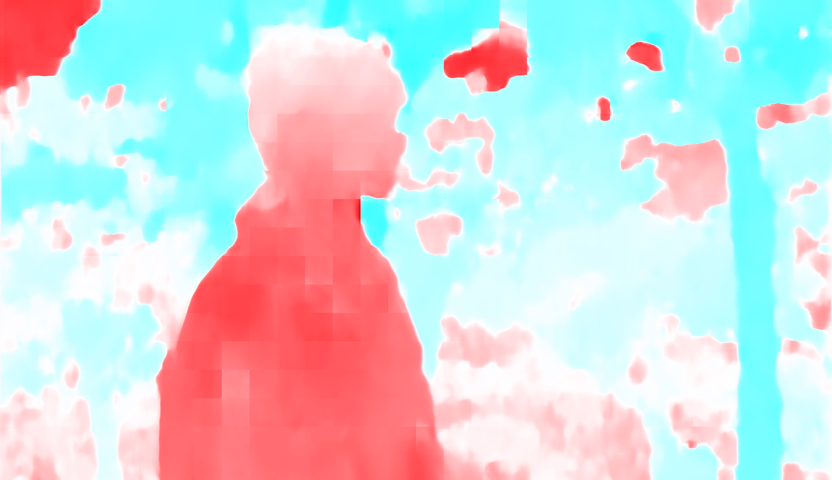}
    \end{minipage}
    \begin{minipage}[b]{0.24\linewidth}
    \centering
    \includegraphics[width=\textwidth]{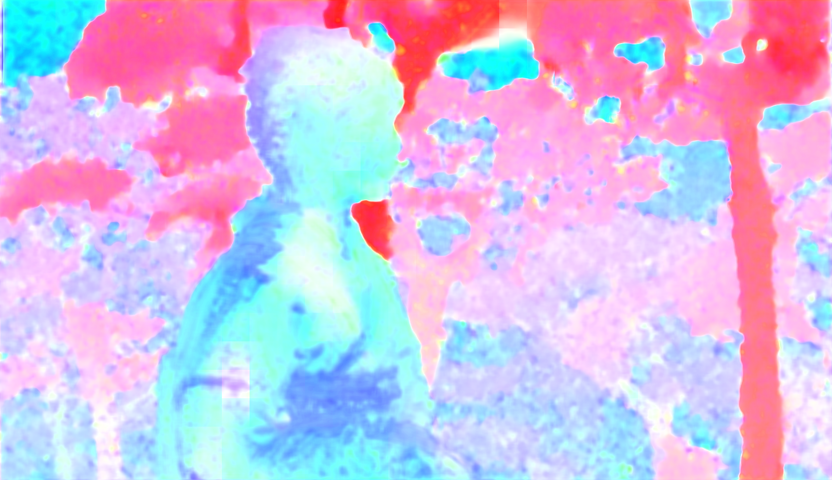}
    \end{minipage}
    \begin{minipage}[b]{0.24\linewidth}
    \centering
    \includegraphics[width=\textwidth]{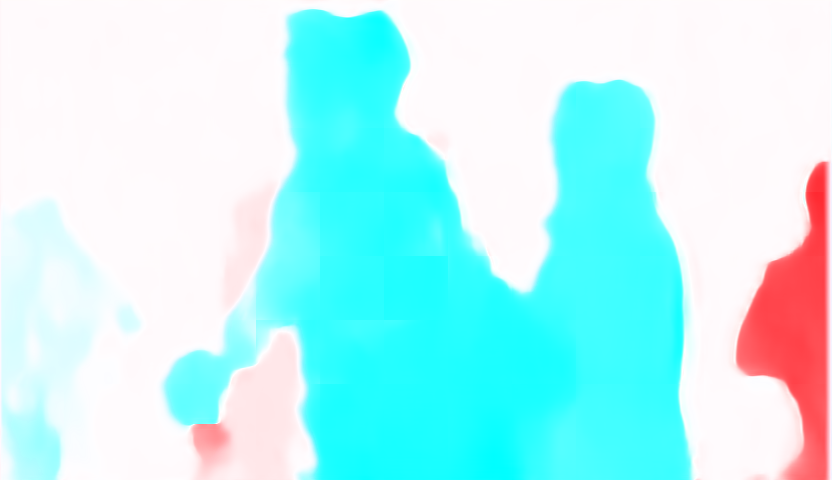}
    \end{minipage}
    \begin{minipage}[b]{0.24\linewidth}
    \centering
    \includegraphics[width=\textwidth]{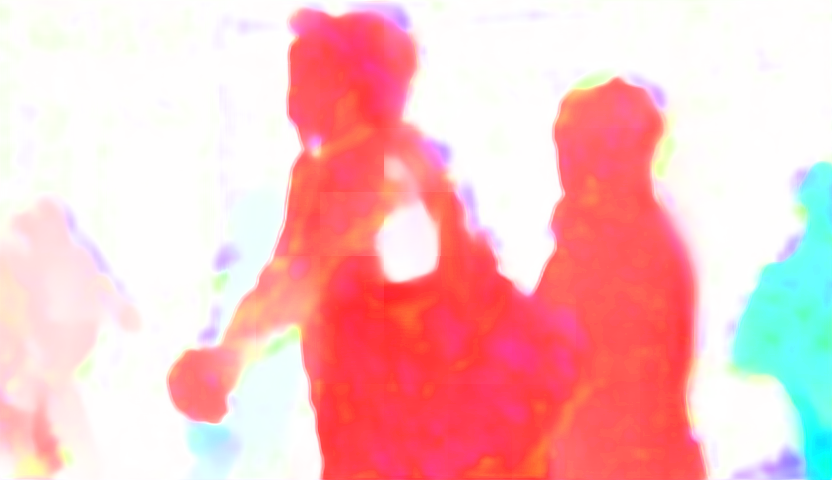}
    \end{minipage}
    \hfill
    \vspace{0.12cm}
    \centerline{\hspace{-0.5cm}(a)\hspace{8.5cm}(b)}
\caption{Visual comparison of motion fields using the learned affine transformation parameters for pairs of frames in (a) Kimono in ClassB and (b) BQMall in ClassC. The left column of (a) and (b) represents the case of bi-directional prediction, while the right column is for uni-directional prediction. The first row shows the residual between the predicted frame and the actual, ground truth target frame.
The next two rows show the predicted motion between an input frame and the target frame.
}
\vspace{-0.2cm}
\label{fig:visual_comp_motion}
\end{figure*}

Fig.~\ref{fig:visual_comp_motion} shows the visualization of the learned affine motion for several pairs of frames from Kimono and BQMall. In the bi-directional case (left column in Fig.~\ref{fig:visual_comp_motion}(a) and (b)), the closest frame on either side of the target frame are used, whereas in the uni-directional case (right column in Fig.~\ref{fig:visual_comp_motion}(a) and (b)), the two closest previous frames are used. The top row shows the residual between the predicted frame and the actual target frame. As expected, bi-directional prediction is more accurate and results in a smaller residual. The next two rows show the estimated motion between an input frame and the target frame. In the left columns of Fig.~\ref{fig:visual_comp_motion}(a) and (b), the two motion fields are oppositely colored, because in the bi-directional prediction case, these motion fields point in the opposite directions. In the right columns of Fig.~\ref{fig:visual_comp_motion}(a) and (b), the two motion fields are similarly colored because in the case of uni-directional prediction, the two fields point in the same direction. We also note that the motion fields capture the boundaries of the moving objects fairly well, further confirming their validity.  

\subsection{Coding performance of two integration methods}
\label{ssec:coding_performance_of_two}
\begin{table*}[t]
\centering
\caption{BD-Bitrate relative to HM-16.20 over three common test conditions}
\label{tbl:overall_coding_performance}
\smallskip\noindent
\resizebox{\linewidth}{!}{%
\setlength\tabcolsep{3.5pt}
\begin{tabular}{@{}ccc|ccccccccc|ccccccccc@{}}
\toprule
\multirow{3}{*}{Class} & \multirow{3}{*}{Sequence} & \multirow{3}{*}{fps} & \multicolumn{9}{c|}{Direct Mode Only (DM)}                                                                                            & \multicolumn{9}{c}{Direct Mode + Reference List Update (DM+RLU)}                                                                         \\ \cmidrule(l){4-21} 
                       &                           &                      & \multicolumn{3}{c|}{LDP}      & \multicolumn{3}{c|}{LD}         & \multicolumn{3}{c|}{RA} & \multicolumn{3}{c|}{LDP}      & \multicolumn{3}{c|}{LD}         & \multicolumn{3}{c}{RA} \\ \cmidrule(l){4-21} 
                       &                           &                      & Y     & U     & \multicolumn{1}{c|}{V}     & Y     & U     & \multicolumn{1}{c|}{V}     & Y           & U           & V          & Y     & U     & \multicolumn{1}{c|}{V}     & Y     & U     & \multicolumn{1}{c|}{V}     & Y          & U           & V           \\ \midrule
\multirow{2}{*}{A}     & PeopleOnStreet            & \multirow{2}{*}{30}  & -8.2  & -10.3 & \multicolumn{1}{c|}{-11.8} & -6.2  & -7.8  & \multicolumn{1}{c|}{-9.7}  & -6.1  & -9.4  & -11.0 & -10.3 & -9.2  & \multicolumn{1}{c|}{-10.7} & -8.0  & -6.9  & \multicolumn{1}{c|}{-8.4}  & -7.7       & -10.7       & -12.6       \\
                       & Traffic                   &                      & -5.8  & -11.0 & \multicolumn{1}{c|}{-11.8} & -4.2  & -9.2  & \multicolumn{1}{c|}{-9.9}  & -3.3  & -4.5  & -4.1  & -6.0  & -10.5 & \multicolumn{1}{c|}{-11.0} & -5.6  & -10.2 & \multicolumn{1}{c|}{-10.9} & -4.9       & -5.5        & -5.0        \\ \midrule
\multirow{5}{*}{B}     & \underbar{BQTerrace}*                 & 60                   & -9.6  & -9.1  & \multicolumn{1}{c|}{-16.3} & -2.8  & -7.9  & \multicolumn{1}{c|}{-16.7} & -2.1  & -1.8  & -3.2  & -8.4  & -1.6  & \multicolumn{1}{c|}{-8.1}  & -2.7  & -4.1  & \multicolumn{1}{c|}{-13.6} & -1.9       & -1.1        & -3.3        \\
                       & BasketballDrive           & \multirow{2}{*}{50}  & -4.3  & -12.3 & \multicolumn{1}{c|}{-12.0} & -1.9  & -8.6  & \multicolumn{1}{c|}{-7.9}  & -2.0  & -4.3  & -3.6  & -6.5  & -13.9 & \multicolumn{1}{c|}{-14.2} & -2.2  & -9.3  & \multicolumn{1}{c|}{-9.7}  & -3.1       & -5.7        & -4.5        \\
                       & \underbar{Cactus}                    &                      & -10.7 & -19.5 & \multicolumn{1}{c|}{-9.4}  & -7.0  & -16.5 & \multicolumn{1}{c|}{-8.6}  & -5.0  & -9.8  & -6.5  & -10.0 & -18.3 & \multicolumn{1}{c|}{-7.5}  & -7.1  & -16.4 & \multicolumn{1}{c|}{-7.9}      & -5.0       & -9.9        & -6.4        \\ 
                       & Kimono                    & \multirow{2}{*}{24}  & -8.9  & -22.2 & \multicolumn{1}{c|}{-4.6}  & -4.8  & -19.3 & \multicolumn{1}{c|}{-3.2}  & -3.2  & -8.0  & -3.6  & -10.8 & -23.3 & \multicolumn{1}{c|}{-1.2}  & -6.4  & -21.1  & \multicolumn{1}{c|}{-0.3}      & -5.3       & -11.2       & -4.5        \\
                       & ParkScene                 &                      & -4.7  & -17.4 & \multicolumn{1}{c|}{-5.5}  & -3.3  & -14.8 & \multicolumn{1}{c|}{-4.4}  & -2.9  & -4.3  & -2.9  & -5.0  & -20.0 & \multicolumn{1}{c|}{-4.2}  & -4.1  & -19.2  & \multicolumn{1}{c|}{-3.2}      & -3.8       & -5.6        & -3.1        \\ \midrule
\multirow{4}{*}{C}     & BQMall                    & 60                   & -9.3  & -19.8 & \multicolumn{1}{c|}{-20.9} & -7.1  & -18.2 & \multicolumn{1}{c|}{-19.3} & -5.8  & -8.6  & -8.6  & -9.6  & -20.3 & \multicolumn{1}{c|}{-21.2} & -7.2  & -18.7 & \multicolumn{1}{c|}{-19.8} & -6.1       & -9.0        & -9.0        \\
                       & BasketballDrill           & \multirow{2}{*}{50}  & -5.5  & -9.3  & \multicolumn{1}{c|}{-9.3}  & -3.8  & -7.3  & \multicolumn{1}{c|}{-7.2}  & -2.1  & -3.8  & -3.5  & -5.7  & -8.9  & \multicolumn{1}{c|}{-8.1}  & -4.6  & -7.7  & \multicolumn{1}{c|}{-7.5}  & -2.2       & -4.1        & -3.7        \\
                       & \underbar{PartyScene}*                &                      & -4.9  & -12.2 & \multicolumn{1}{c|}{-11.7} & -3.1  & -10.2 & \multicolumn{1}{c|}{-9.7}  & -4.5  & -5.9  & -5.4  & -3.0  & -10.9 & \multicolumn{1}{c|}{-9.8}  & -3.0  & -10.8 & \multicolumn{1}{c|}{-10.2} & -4.3       & -6.1        & -5.6        \\
                       & RaceHorsesC               & 30                   & -1.9  & -4.3  & \multicolumn{1}{c|}{-5.8}  & -1.3  & -3.2  & \multicolumn{1}{c|}{-4.6}  & -1.1  & -2.0  & -2.5  & -2.6  & -4.5  & \multicolumn{1}{c|}{-6.8}  & -1.9  & -4.0  & \multicolumn{1}{c|}{-6.2}  & -1.6       & -3.0        & -4.3        \\ \midrule
\multirow{4}{*}{D}     & \underbar{BQSquare}*                  & 60                   & -6.8  & -4.1  & \multicolumn{1}{c|}{-11.3} & -3.7  & -6.8  & \multicolumn{1}{c|}{-13.4} & -4.4  & -4.0  & -5.5  & -1.2  & 5.6   & \multicolumn{1}{c|}{-3.1}  & -3.3  & -1.1  & \multicolumn{1}{c|}{-9.9}  & -3.5       & -4.2        & -5.9        \\
                       & BasketballPass            & \multirow{2}{*}{50}  & -5.9  & -10.5 & \multicolumn{1}{c|}{-8.6}  & -4.5  & -8.7  & \multicolumn{1}{c|}{-7.2}  & -4.5  & -6.3  & -5.2  & -7.6  & -11.8 & \multicolumn{1}{c|}{-10.0} & -5.4  & -9.8  & \multicolumn{1}{c|}{-7.8}  & -5.4       & -7.2        & -5.6        \\
                       & \underbar{BlowingBubbles}            &                      & -5.9  & -13.0 & \multicolumn{1}{c|}{-12.5} & -4.4  & -11.7 & \multicolumn{1}{c|}{-12.0} & -4.5  & -5.6  & -5.8  & -4.8  & -11.5 & \multicolumn{1}{c|}{-10.9} & -4.3  & -12.1 & \multicolumn{1}{c|}{-11.5} & -4.7       & -5.8        & -6.0        \\
                       & RaceHorses                & 30                   & -2.8  & -5.6  & \multicolumn{1}{c|}{-6.9}  & -2.2  & -5.3  & \multicolumn{1}{c|}{-6.1}  & -2.5  & -4.1  & -4.4  & -3.5  & -6.3  & \multicolumn{1}{c|}{-7.3}  & -2.9  & -5.7  & \multicolumn{1}{c|}{-6.7}  & -3.6       & -5.3        & -5.8        \\ \midrule
\multirow{3}{*}{E}     & FourPeople                & \multirow{3}{*}{60}  & -12.2 & -13.1 & \multicolumn{1}{c|}{-17.2} & -10.4 & -11.2 & \multicolumn{1}{c|}{-15.5} & \multicolumn{3}{c|}{--}         & -12.9 & -12.0 & \multicolumn{1}{c|}{-17.3} & -10.6 & -10.7 & \multicolumn{1}{c|}{-15.2} & \multicolumn{3}{c}{--}          \\
                       & \underbar{Johnny}*                    &                      & -11.3 & -9.2  & \multicolumn{1}{c|}{-6.9}  & -8.6  & -7.6  & \multicolumn{1}{c|}{-6.3}  & \multicolumn{3}{c|}{--}         & -10.4 & -3.7  & \multicolumn{1}{c|}{-1.5}  & -8.1  & -3.1  & \multicolumn{1}{c|}{-1.8}  & \multicolumn{3}{c}{--}          \\
                       & KristenAndSara            &                      & -12.4 & -12.0 & \multicolumn{1}{c|}{-14.4} & -10.1 & -10.3 & \multicolumn{1}{c|}{-13.1} & \multicolumn{3}{c|}{--}         & -12.7 & -9.5  & \multicolumn{1}{c|}{-13.3} & -10.4 & -8.3  & \multicolumn{1}{c|}{-10.9} & \multicolumn{3}{c}{--}          \\ \midrule
\multicolumn{3}{c|}{Average}                                               & -7.3  & -12.0 & \multicolumn{1}{c|}{-10.9} & -5.0  & -10.3 & \multicolumn{1}{c|}{-9.7}  & -3.6  & -5.5  & -5.1  & -7.3  & -10.6 & \multicolumn{1}{c|}{-9.2}  &  -5.4  & -10.0  & \multicolumn{1}{c|}{-9.0}      & -4.2       & -6.3        & -5.7        \\ \bottomrule
\end{tabular}}
\end{table*}

\begin{figure}[t]
    \centering
    \begin{minipage}[b]{1.0\linewidth}
    \centering
    \includegraphics[width=\textwidth]{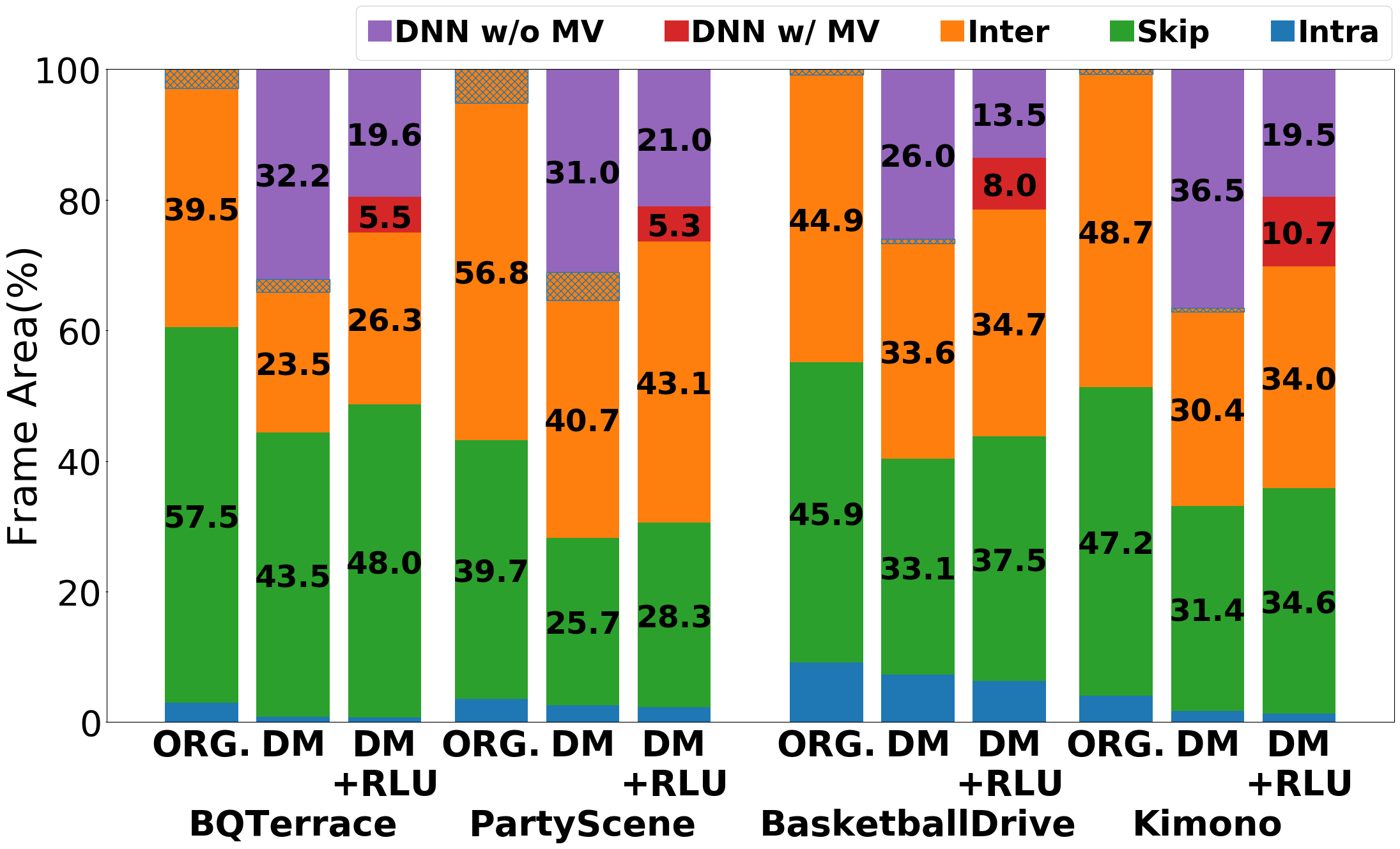}
    \centerline{(a)}
    \end{minipage}
    
    \vspace{.1cm}
    \centering
    \begin{minipage}[b]{1.0\linewidth}
    \centering
    \includegraphics[width=\textwidth]{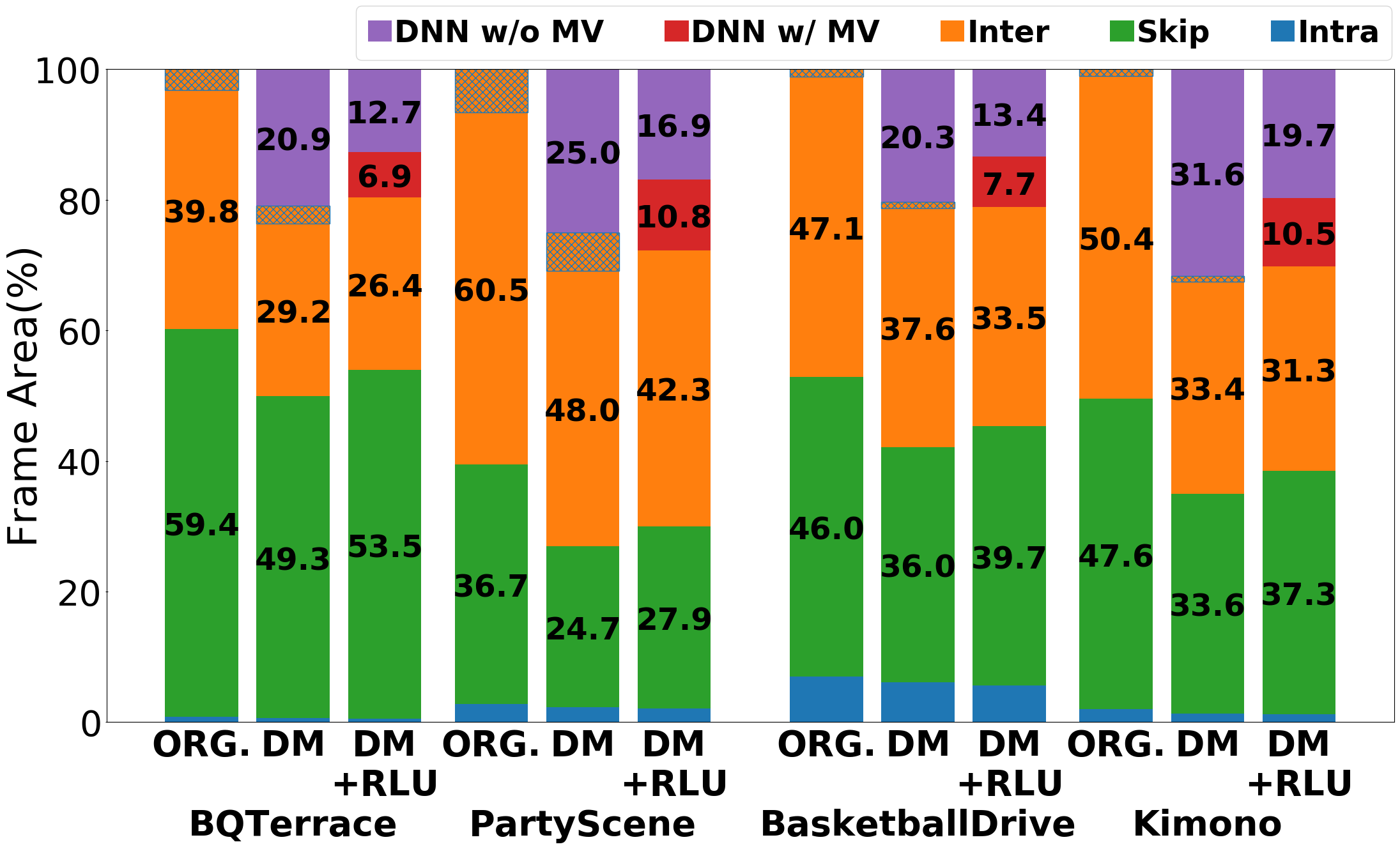}
    \centerline{(b)}
    \end{minipage}

    \vspace{.1cm}
    \centering
    \begin{minipage}[b]{1.0\linewidth}
    \centering
    \includegraphics[width=\textwidth]{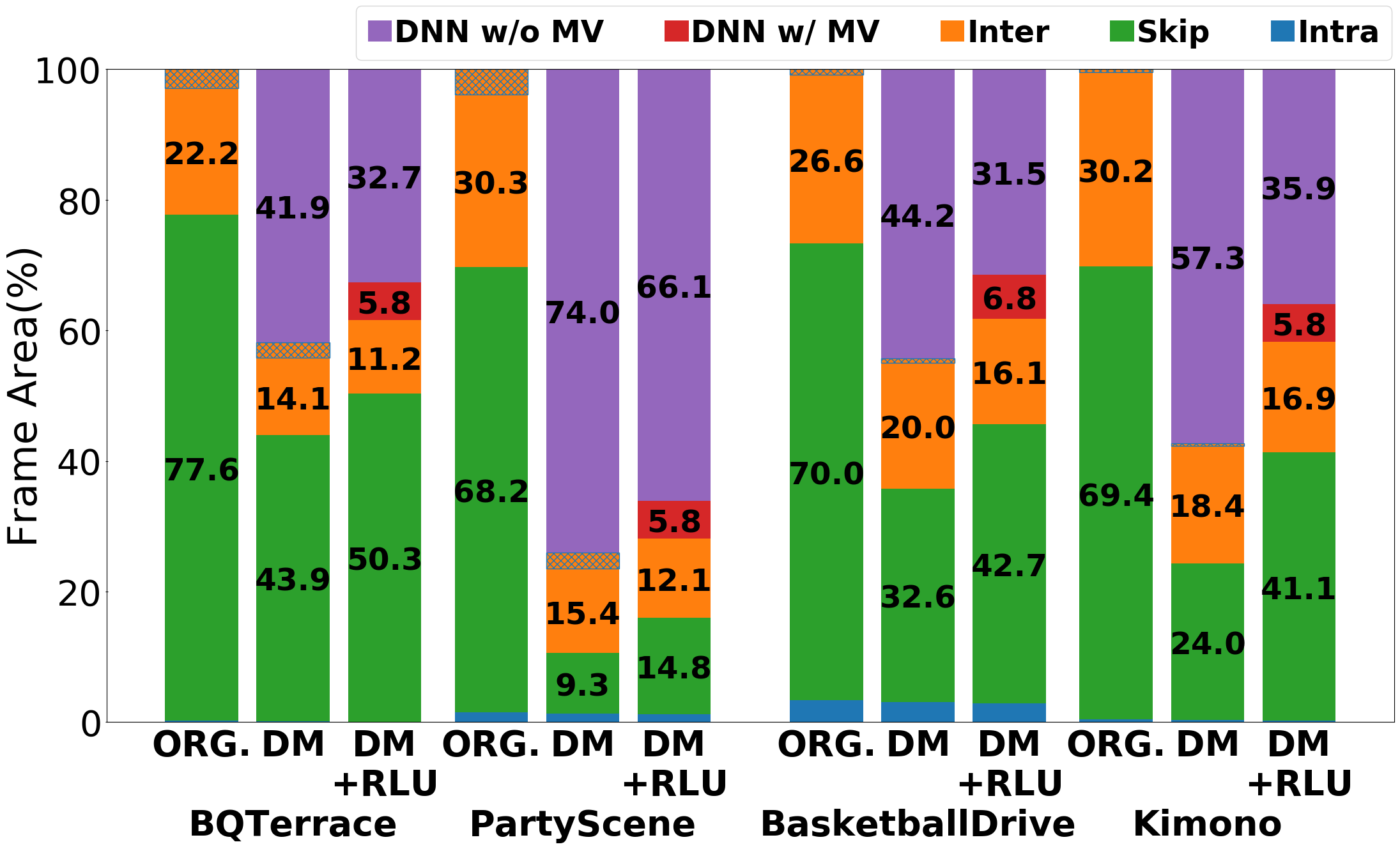}
    \centerline{(c)}
    \end{minipage}
\caption{Frame area of selected modes for various sequences in the three coding configuration: (a) LDP, (b) LD, and (c) RA}
\label{fig:comp_r3_ref_analysis}
\end{figure}

Table~\ref{tbl:overall_coding_performance} shows the BD-Bitrate~\cite{bd_br} of the proposed methods integrated into HM-16.20, where HM-16.20 itself is used as the anchor. Two integration methods are examined: ``Direct'' mode (DM) by itself, and DM combined with reference list update (RLU) (DM+RLU). As shown in the table, both methods achieve considerable bit reduction compared to HM-16.20, ranging from -3.6\% to -12.0\%, depending on the color channel and coding configuration. 
Since DM+RLU allows variable-size prediction blocks to be used, it can be expected that this method will usually have better coding performance than DM by itself. The results in the table confirm this intuition. 
Indeed, for LD and RA configurations, DM achieves -5.0\% and -3.6\% bit savings on average, respectively, while DM+RLU achieves additional -0.4\% and -0.6\% bit savings, respectively. 

Bit savings are especially large in sequences that include camera panning, relatively large motion, and/or few moving objects, implying that in such cases, the affine motion-based network captures the motion well. 
In some sequences, bit savings of about -12\% in LDP configuration for Y-channel are achieved, for both DM and DM+RLU. Although DM+RLU shows better performance than DM on average, there are six sequences (underlined in the table) where DM achieves better coding efficiency. In four of these sequences, indicated with asterisk (*), DM has better performance across all tested configurations. 
Motion characteristics within these sequences include zoom-in/out, complex background texture and some occlusion/disocclusion. Based on the results, it appears that under these motion characteristics, the replaced frame may still contain some useful information that is not present in the DNN-generated frame, so its replacement in the reference list does not lead to better coding efficiency, despite the added benefit of variable block-shape prediction. 

Fig.~\ref{fig:comp_r3_ref_analysis} shows the percentage of frame area taken up by selected coding modes in the four sequences across all four tested QP values. In two of these sequences (BQTerrace and PartyScene)  DM has better performance than DM+RLU, while in the other two (BasketballDrive and Kimono) DM+RLU has better performance. For each sequence, three bars are shown: one corresponding to HM-16.20 by itself (``ORG''), one for DM, and one for DM+RLU. Included in the figure are three conventional coding modes (Intra, Inter, Skip) and two coding modes associated with the proposed methods: ``DNN w/o MV'' and ``DNN w/ MV.'' The first of these, ``DNN w/o MV,'' means that the DNN-generated frame is selected as a predictor for the corresponding block without an additional MV; this mode is available in both DM and DM+RLU. The second mode, ``DNN w/ MV,'' means that the DNN-generated frame is selected as a reference for the corresponding block with an additional MV; this mode is only available in DM+RLU. 

From the results, we see that in the case of DM, ``DNN w/o MV'' takes over not only a significant portion of area previously associated with Inter prediction, but also some part of the area where Skip mode was used. This is especially noticeable in Fig.~\ref{fig:comp_r3_ref_analysis}(c), which corresponds to the RA coding configuration. The same trend is visible in the case of DM+RLU, although here there are two modes that use the DNN-predicted frame. The ``DNN w/ MV'' mode uses additional motion-related bits to signal the MV, and it is selected less frequently than the ``DNN w/o MV'' mode. 

The small rectangular areas filled with blue patterns in the orange bars indicate Inter blocks that refer to the frame that is replaced by the DNN-generated frame in DM+RLU. The fact that these rectangles are smaller than the red rectangles in DM+RLU bars implies that DNN-generated frame ends up being selected more frequently than the frame it has replaced in the reference list. This validates the effectiveness of the reference frame replacement strategy presented in Section~\ref{sec:integration}.

Another observation from Fig.~\ref{fig:comp_r3_ref_analysis} is that in BQTerrace and PartyScene (the sequences where DM outperformed DM+RLU), the reference frame that was replaced by the DNN-generated frame in RLU was used more frequently than in BasketballDrive and Kimono (the sequences where DM+RLU outperformed DM). Again, this is intuitively clear, and sheds light on when DM+RLU might be a better strategy that DM by itself. If the reference frame that is replaced by the DNN-generated frame is used relatively frequently even in the presence of DM (as is the case in BQTerrace and PartyScene), this means that it contains useful information that may not be present in the DNN-generated frame; in such cases, its replacement may not be warranted. On the other hand, if the reference frame is not used frequently in the presence of DM (as is the case in BasketballDrive and Kimono), this means it does not have much useful information to offer beyond what is already in the DNN-generated frame, so its replacement by the DNN-generated frame would improve the coding performance.

\begin{table}[t]
\centering
\caption{Comparison of average run-time per frame}
\label{tbl:comp_complexity}
\smallskip\noindent
\resizebox{\linewidth}{!}{%
\setlength\tabcolsep{5pt}
\begin{tabular}{@{}c|c|p{0.05\textwidth}*{4}{>{\centering\arraybackslash}p{0.05\textwidth}}@{}}
\toprule
\multicolumn{2}{c|}{OS}                                                                             & \multicolumn{4}{c}{Ubuntu 16.04.6 LTS}                  \\ \midrule
\multicolumn{2}{c|}{GPU}                                                                   & \multicolumn{4}{c}{1 $\times$ GeForce RTX 2080 Ti with 11 GiB} \\ \midrule
\multirow{2}{*}{DNNs} & \multirow{2}{*}{\begin{tabular}[c]{@{}c@{}}Size \\ (GiB)\end{tabular}} & \multicolumn{4}{c}{Average run-time (sec)}                 \\ \cmidrule(l){3-6} 
    &       & 1080P     & 720P      & 832$\times$480       & 416$\times$240      \\ \midrule
{~\cite{deep_frame_prediction}}     & 0.646     & 0.476     & 0.214     & 0.103     & 0.032        \\
Ours                                & \textbf{0.164}    & \textbf{0.162}    & \textbf{0.064}    & \textbf{0.033}     & \textbf{0.016} \\ \bottomrule
\end{tabular}}
\end{table}

\subsection{Run-time complexity}

In this section we analyze the run-time complexity of the proposed model and compare it with our previous work~\cite{deep_frame_prediction}, which used a DNN architecture similar to~\cite{hevc_with_sep_conv_for_ra, xia2019deep}.\footnote{Other DNNs~\cite{hevc_with_sep_conv_for_ra, xia2019deep, lee2020deep} are not publicly available.} The first two rows in Table~\ref{tbl:comp_complexity} show the operating system and GPU information used for this experiment. The second column represents the size of the memory required to hold each DNN in the GPU. Our proposed network requires about four times less memory than the previous model~\cite{deep_frame_prediction}, which is not surprising considering the correspondingly smaller number of parameters. For the run-time comparison, all experiments are done on a single GPU. The experiments were based on several sequences with different resolutions. For each resolution, indicated in the fourth row, run-time was averaged over 100 frames. The smaller size of our model contributes to its faster operation, as shown in the last two rows of the table, where better results are indicated in bold. Depending on the resolution, the run-time of the proposed model is 2-3 times shorter than that of~\cite{deep_frame_prediction}.


\subsection{Visualization of blocks coded using the proposed methods}


\begin{figure*}[t]
    \centering
    \begin{minipage}[b]{0.48\linewidth}
    \centering
    \includegraphics[width=\textwidth]{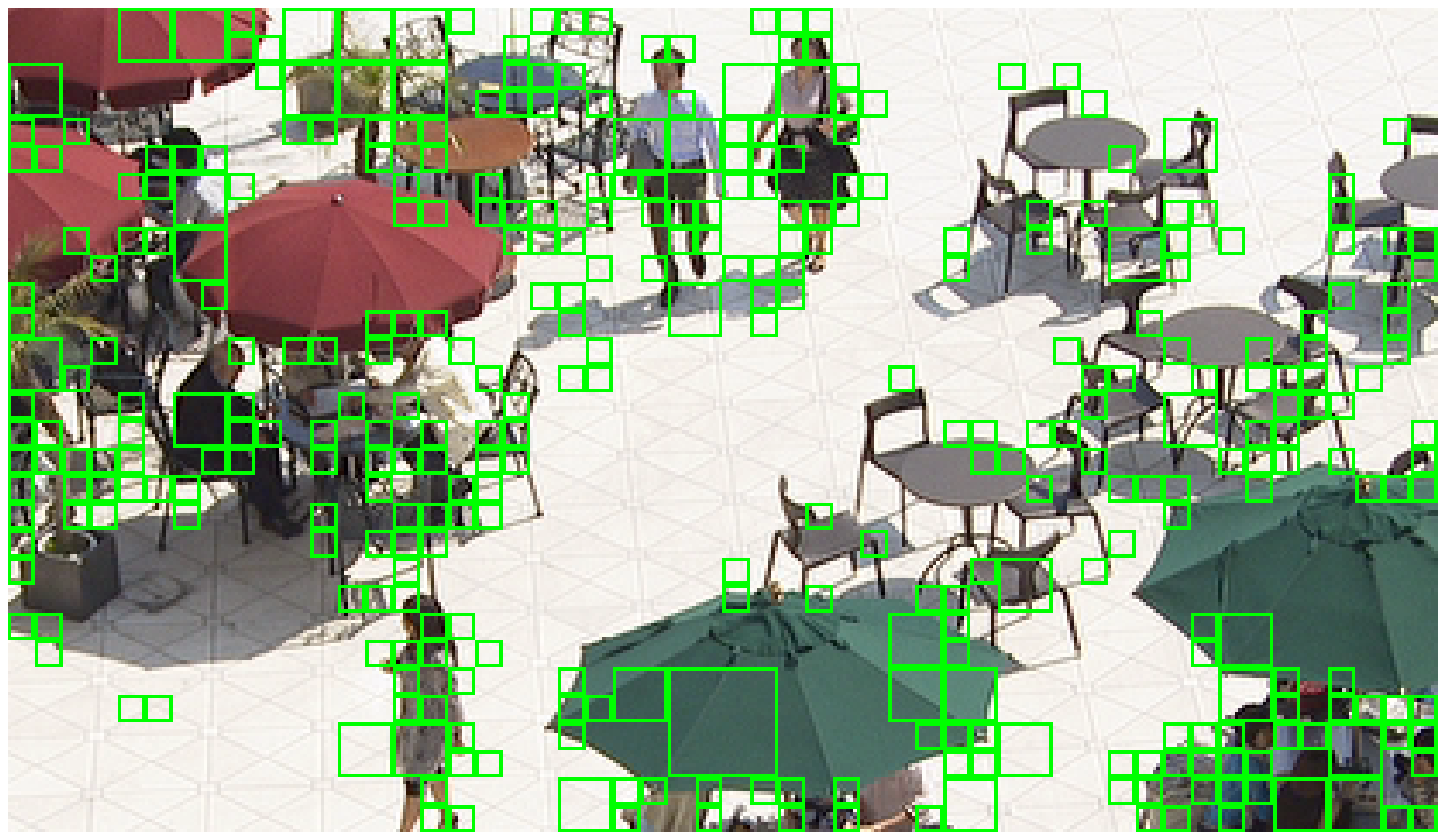}
    \centerline{(a)}
    \end{minipage}
    \begin{minipage}[b]{0.48\linewidth}
    \centering
    \includegraphics[width=\textwidth]{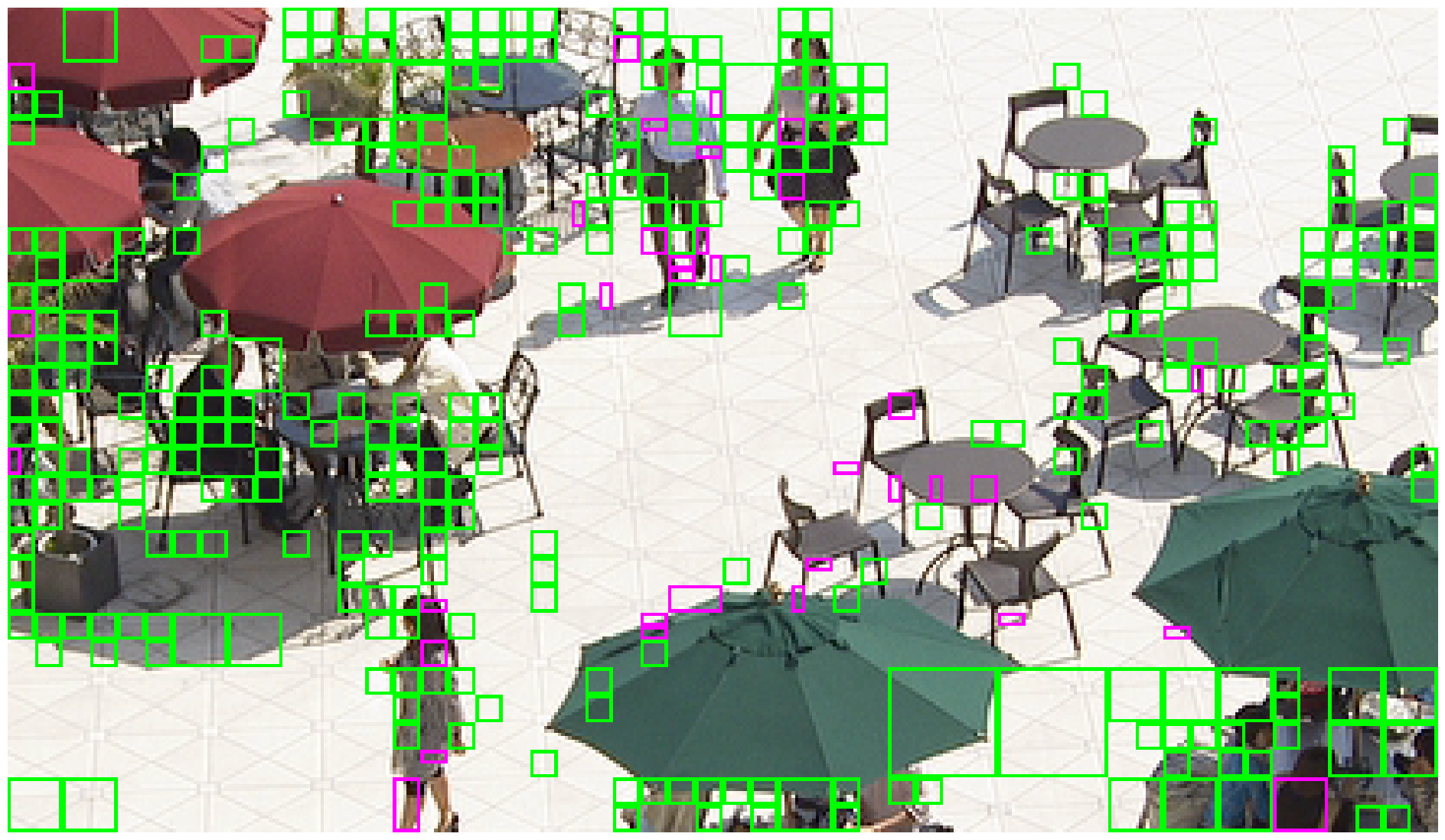}
    \centerline{(b)}
    \end{minipage}
    \hfill
    
    \vspace{0.08cm}
    \begin{minipage}[b]{0.48\linewidth}
    \centering
    \includegraphics[width=\textwidth]{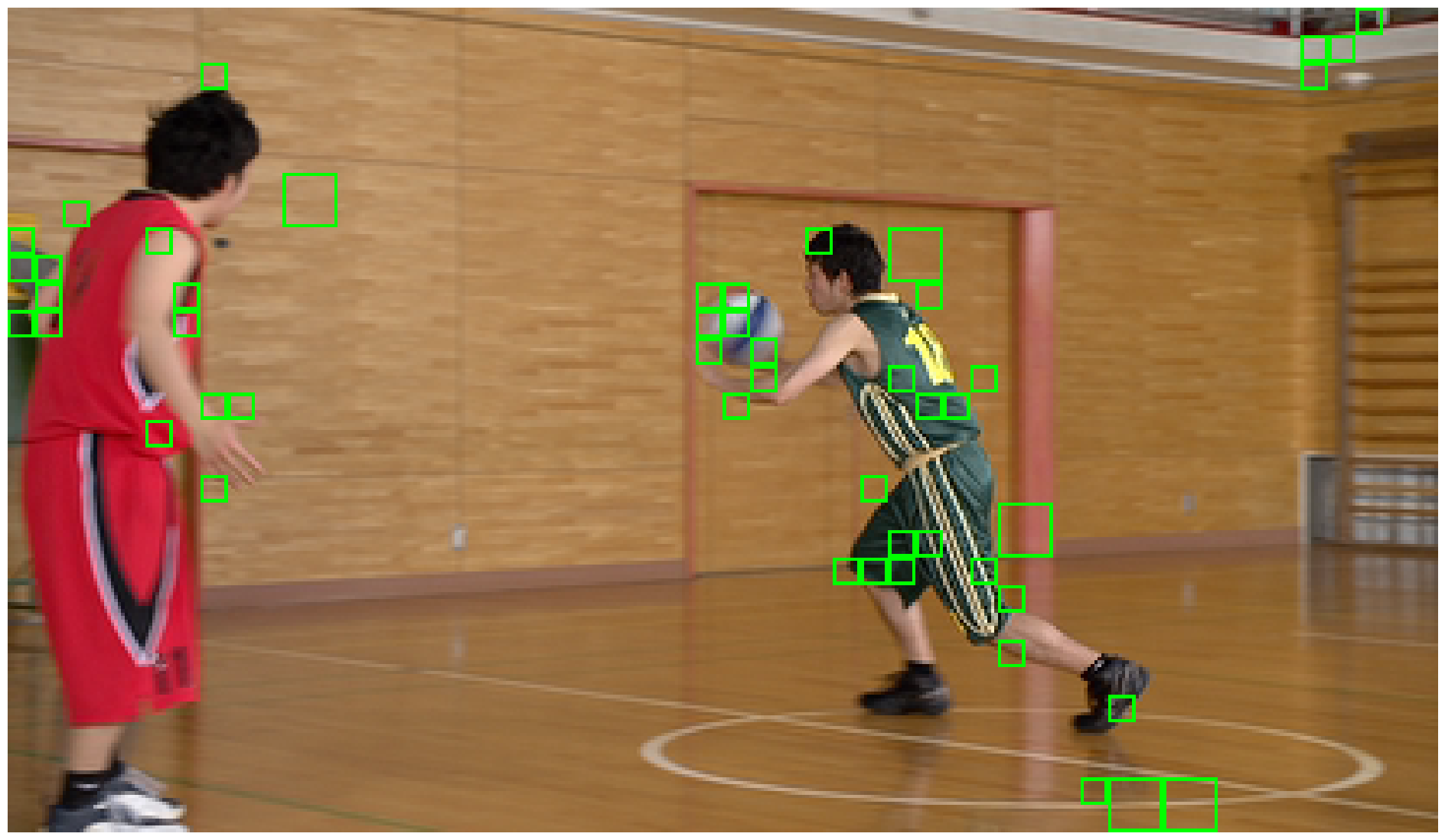}
    \centerline{(c)}
    \end{minipage}
    \begin{minipage}[b]{0.48\linewidth}
    \centering
    \includegraphics[width=\textwidth]{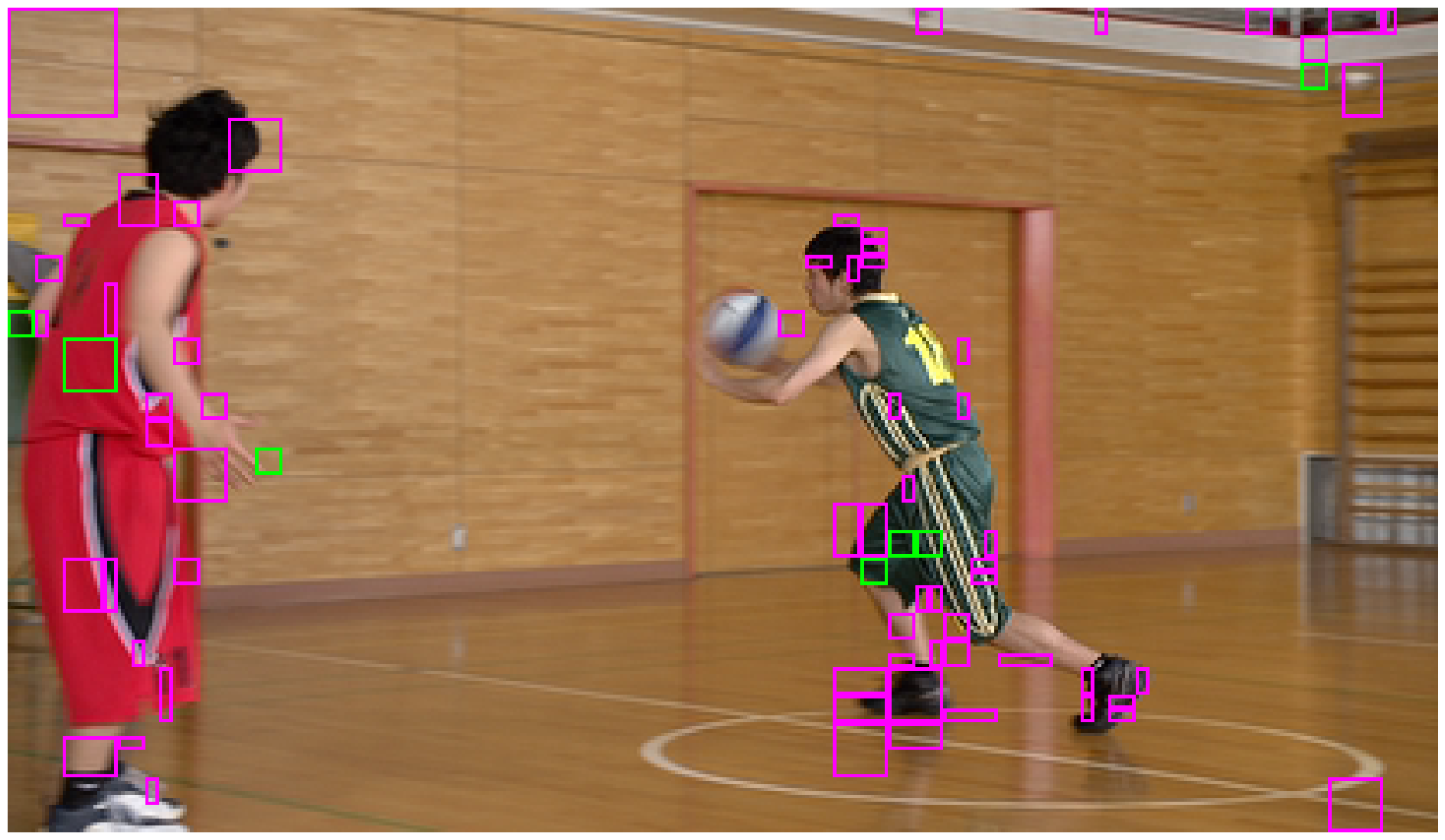}
    \centerline{(d)}
    \end{minipage}
    \hfill
\caption{Visualized blocks coded with the proposed methods with QP=22 in the LDP configuration. The top row shows frame 91 of BQSquare and the bottom row shows frame 53 of BasketballPass. 
The left column corresponds to the DM configuration and the right column is for DM+RLU. Blocks coded with the ``DNN w/o MV'' mode are outlined in green, and those coded with the ``DNN w/ MV'' mode are outlined in magenta color. 
}
\label{fig:visualized_selected_region}
\end{figure*}

Fig.~\ref{fig:visualized_selected_region} visualizes blocks coded with the proposed methods. We selected sequences where there was a relatively large gap in coding efficiency between DM and DM+RLU. Specifically, we chose BQSquare and BasketballPass coded with QP=22 in the LDP configuration. 
These sequences have similar characteristic as BQTerrace and BasketballDrive, respectively, but with lower frame resolution. Fig.~\ref{fig:visualized_selected_region}(a) and (c) correspond to the DM case, and Fig.~\ref{fig:visualized_selected_region}(b) and (d) are for the DM+RLU case. Blocks coded with the ``DNN w/o MV'' mode are outlined in green, and those coded with the ``DNN w/ MV'' mode are outlined in magenta color. 

In Fig.~\ref{fig:visualized_selected_region}(a), around 31\% of the frame area is coded with the ``DNN w/o MV'' mode. Specifically, areas near object edges seem to be frequently selected for coding using the ``DNN w/o MV'' mode, indicating that in such cases, the DNN-generated frame without additional motion information is able to offer improved coding efficiency. 
Looking at Fig.~\ref{fig:visualized_selected_region}(b), which shows the same frame coded using DM+RLU, we see that relatively few blocks end up coded with the ``DNN w/ MV'' mode, and most are still coded with the ``DNN w/o MV'' mode. This indicates that the affine motion model captures motion characteristics in BQSquare fairly well, and little useful information can be brought by additional motion information.

For the sequence BasketballPass, only about 4\% of the frame is coded with the ``DNN w/o MV'' mode, as shown in Fig.~\ref{fig:visualized_selected_region}(c). However, these blocks are still around object boundaries, as was the case with BQSquare. As shown in Fig.~\ref{fig:visualized_selected_region}(d), once additional motion information is allowed through the ``DNN w/ MV'' mode, this becomes the more efficient option, and most blocks that do refer to the DNN-generated frame now use this mode. Clearly, the type of motion present around object boundaries in BasketballPass is not sufficiently well modeled by an affine-based DNN model by itself, so coding efficiency does benefit from additional motion information allowed in the ``DNN w/ MV'' mode.  



\subsection{Performance comparison with related methods}

\begin{table}[t]
\centering
\caption{Luminance BD-Bitrate of our new methods and our previous work~\cite{deep_frame_prediction} relative to HM-16.20}
\label{tbl:compare_with_prev}
\smallskip\noindent
\resizebox{\linewidth}{!}{%
\setlength\tabcolsep{2.5pt}
\begin{tabular}{c|ccc|ccc|ccc}
\toprule
\multirow{3}{*}{Class} & \multicolumn{3}{c|}{LDP}                                 & \multicolumn{3}{c|}{LD}                               & \multicolumn{3}{c}{RA}                                                 \\ \cmidrule(l){2-10} 
                       & \multirow{2}{*}{~\cite{deep_frame_prediction}} & \multicolumn{2}{c|}{Ours}       & \multirow{2}{*}{~\cite{deep_frame_prediction}} & \multicolumn{2}{c|}{Ours}     & {\multirow{2}{*}{~\cite{deep_frame_prediction}}} & \multicolumn{2}{c}{Ours} \\ \cmidrule(l){3-4} \cmidrule(l){6-7} \cmidrule(l){9-10} 
                       &                       & DM             & \begin{tabular}[c]{@{}c@{}}DM\\ +RLU\end{tabular}         &                       & DM            & \begin{tabular}[c]{@{}c@{}}DM\\ +RLU\end{tabular}        &  & DM      & \begin{tabular}[c]{@{}c@{}}DM\\ +RLU\end{tabular}          \\  \midrule
A                      & -4.7                  & -7.0           & \textbf{-8.2}  & -2.3                  & -5.2          & \textbf{-6.8} & \multicolumn{1}{c}{-3.2}                  & -4.7    & \textbf{-6.3}  \\
B                      & -4.5                  & -7.6           & \textbf{-8.1}  & -2.1                  & -4.0          & \textbf{-4.5} & \multicolumn{1}{c}{-1.8}                  & -3.0    & \textbf{-3.8}  \\
C                      & -3.3                  & \textbf{-5.4}           & -5.2  & -2.2                  & -3.8          & \textbf{-4.2} & \multicolumn{1}{c}{-2.7}                  & -3.4    & \textbf{-3.5}  \\
D                      & -3.4                  & \textbf{-5.3}  & -4.3           & -2.2                  & -3.7          & \textbf{-4.0} & \multicolumn{1}{c}{-3.3}                  & -4.0    & \textbf{-4.3}  \\
E                      & -9.1                  & \textbf{-12.0} & \textbf{-12.0} & -5.8                  & \textbf{-9.7} & \textbf{-9.7} & \multicolumn{3}{c}{--}                                               \\ \midrule
Average                & -4.8                  & \textbf{-7.3}           & \textbf{-7.3}           & -2.9                  & -5.0          & \textbf{-5.4}          & -2.6   &  -3.6   & \textbf{-4.2} \\ \bottomrule 
\end{tabular}}
\end{table}

We first compare our new methods with our previous work~\cite{deep_frame_prediction} in terms of coding efficiency. 
The DNNs of both works are integrated with HM-16.20. Table~\ref{tbl:compare_with_prev} shows the average luminance BD-Bitrate within each class of test sequences, against the default HM-16.20. 
In all configurations, our new methods provide additional bit savings of -1.2\% to -2.5\%, compared to our previous work~\cite{deep_frame_prediction}.  
As mentioned in Section~\ref{ssec:coding_performance_of_two}, DM+RLU is better than DM on average, but there are cases where DM outperforms DM+RLU, especially in the LDP configuration. 

\begin{table}[t]
\centering
\caption{Luminance BD-Bitrate relative to HM-16.6 with RA configuration for higher QPs}
\label{tbl:bdrate_ra}
\smallskip\noindent
\resizebox{\linewidth}{!}{%
\setlength\tabcolsep{3.0pt}
\renewcommand{\arraystretch}{0.8}
\begin{tabular}{@{}cccccc@{}}
\toprule
\multicolumn{2}{c}{\# of trainable param.}                 & 21.68M                      & 21.67M                & 27.84M          & \textbf{5.5M}         \\ \midrule
Class                  & Sequence       & \begin{tabular}[c]{@{}c@{}}
\cite{hevc_with_sep_conv_for_ra}\end{tabular} & \begin{tabular}[c]{@{}c@{}}
\cite{deep_frame_prediction}\end{tabular} & \begin{tabular}[c]{@{}c@{}}
\cite{xia2019deep}\end{tabular} & Ours \\ \midrule
\multirow{6}{*}{B}     & BasketballDrive              & -1.1                         & -1.5           & -                           & \textbf{-3.6}     \\
                       & BQTerrace                    & -0.2                         & \textbf{-1.0}  & -                           & -0.5     \\
                       & Cactus                       & -4.6                         & -3.1           & -                           & \textbf{-5.7}     \\
                       & Kimono                       & -1.7                         & -0.9           & -                           & \textbf{-6.1}     \\
                       & ParkScene                    & -2.6                         & -2.9           & -                           & \textbf{-6.1}     \\ \cmidrule(l){2-6} 
                       & \multicolumn{1}{c|}{Average} & -2.0                         & -1.9           & -                           & \textbf{-4.4}     \\ \midrule
\multirow{5}{*}{C}     & BasketballDrill              & -3.2                         & -2.4           & \textbf{-4.8}               & -4.0     \\
                       & BQMall                       & -6.0                         & -7.1           & \textbf{-8.4}               & -7.8    \\
                       & PartyScene                   & -3.0                         & -4.3           & \textbf{-4.8}               & \textbf{-4.8}     \\
                       & RaceHorses                   & -0.8                         & -1.4           & -2.4                        & \textbf{-3.6}     \\ \cmidrule(l){2-6} 
                       & \multicolumn{1}{c|}{Average} & -3.2                         & -3.8           & \textbf{-5.1}               & -5.0     \\ \midrule
\multirow{5}{*}{D}     & BasketballPass               & -5.4                         & -6.7           & -9.5                        & \textbf{-9.6}     \\
                       & BQSquare                     & -7.1                         & -4.7           & \textbf{-7.3}               & -5.6     \\
                       & BlowingBubbles               & -4.1                         & -4.3           & -5.1                        & \textbf{-5.6}     \\
                       & RaceHorses                   & -2.2                         & -2.9           & -5.9                        & \textbf{-6.4}     \\ \cmidrule(l){2-6} 
                       & \multicolumn{1}{c|}{Average} & -4.7                         & -4.6           & \textbf{-7.0}               & -6.8     \\ \midrule
\multicolumn{2}{c|}{Average}                          & -3.2                         & -3.3           & -                           & \textbf{-5.3}     \\ \bottomrule
\end{tabular}}
\end{table}

Next, in order to compare the performance with other works, we integrated our methods into HM-16.6 and measured the performance according to the test conditions used in these earlier works. With reference to~\cite{hevc_with_sep_conv_for_ra, deep_frame_prediction, xia2019deep}, only the initial two seconds of each sequence 
are coded with slightly higher QPs than usual, QP $\in\{27, 32, 37, 42\}$, in the RA configuration. In Table~\ref{tbl:bdrate_ra}, ``Ours'' refers to our DM+RLU method. 
The table shows the luminance BD-Bitrate of the four methods across the sequences that were reported in previous works. 
Additionally, the size of each network is shown in the first row. Compared to the previous works~\cite{hevc_with_sep_conv_for_ra, deep_frame_prediction, xia2019deep}, our network is significantly smaller, around quarter of the size of the previous models. Despite the small size, our average coding gain is superior to~\cite{hevc_with_sep_conv_for_ra, deep_frame_prediction}. Compared to the latest work~\cite{xia2019deep}, our method achieves better coding efficiency in four out of eight sequences tested in~\cite{xia2019deep}, and the same performance on one other sequence (PartyScene). However, the average coding gain is slightly lower than~\cite{xia2019deep} due to the strong performance of that method in the remaining three sequences. Nonetheless, considering the performance across all sequences, especially in the context of model size, we believe it is justified to say our model is competitive with~\cite{xia2019deep}. 

\begin{table}[t]
\large
\centering
\caption{Luminance BD-Bitrate relative to HEVC over various common test conditions}
\label{tbl:ldp_ld_work}
\smallskip\noindent
\resizebox{\linewidth}{!}{%
\begin{tabular}{@{}c|cc|ccc|ccc@{}}
\toprule
\begin{tabular}[c]{@{}c@{}}\# of \\ trainable\\  param.\end{tabular} & \multicolumn{2}{c|}{\begin{tabular}[c]{@{}c@{}}0.58M $\times$ 4 (QPs) \\ = 2.32M\end{tabular}}            & \multicolumn{3}{c|}{43.36M}  & \multicolumn{3}{c}{5.5M}      \\ \midrule
\multirow{2}{*}{Class}                                               & \multicolumn{2}{c|}{\begin{tabular}[c]{@{}c@{}}Huo \textit{et al.}~\cite{extrapolation_with_reference_alignment}\\ (w\textbackslash HM-12.0)\end{tabular}} & \multicolumn{3}{c|}{\begin{tabular}[c]{@{}c@{}}Lee \textit{et al.}~\cite{lee2020deep}\\ (w\textbackslash HM-16.6)\end{tabular}} & \multicolumn{3}{c}{\begin{tabular}[c]{@{}c@{}}Ours\\ (w\textbackslash HM-16.6)\end{tabular}} \\ \cmidrule(l){2-9} 
    & LDP           & LD            & RA            & LDP           & LD            & RA            & LDP        & LD       \\ \midrule
B   & -7.8          & -4.0          & -2.2          & \textbf{-8.5} & \textbf{-5.2} & \textbf{-3.7} & -7.8       & -4.1     \\
C   & -4.6          & -2.6          & -3.0          & -4.7          & \textbf{-4.0} & \textbf{-3.5} & \textbf{-5.5} & -3.9  \\
D   & -3.6          & -2.4          & \textbf{-4.3} & -4.1          & \textbf{-3.9} & \textbf{-4.3} & \textbf{-5.4} & -3.8  \\
E   & -12.2         & -8.7          & -1.4          & \textbf{-12.8}& \textbf{-10.5}& \textbf{-4.5} & -12.5         & -10.1 \\ \midrule
Average & -6.7      & -4.1          & -2.8          & \textbf{-7.5} & \textbf{-5.9} & \textbf{-3.9} & \textbf{-7.5} & -5.1  \\ \bottomrule
\end{tabular}}
\end{table}

%

Additionally, the coding performance of our method is compared with other recent works focusing on extrapolation~\cite{extrapolation_with_reference_alignment} and the multi-resolution frame estimation 
DNN~\cite{lee2020deep}. The extrapolation networks in~\cite{extrapolation_with_reference_alignment} are integrated with HM-12.0, but
we believe the comparison with our method integrated into HM-16.6 is valid, because both version of HM use the same parameters associated with the reference lists in the LD and LDP configuration. We confirmed this by comparing the RD performance of HM-16.6 and HM-12.0, and found the difference in coding efficiency to be negligible. Specifically, the luminance BD-Bitrate of HM-12.0 against HM-16.6 was about 0.02\% and 0.05\%, averaged across all sequences from Class B to E, in LD and LDP configurations, respectively. 
Therefore, by following the experimental setup in~\cite{extrapolation_with_reference_alignment}, we evaluate the proposed DM+RLU method (``Ours'' in Table~\ref{tbl:ldp_ld_work}) integrated with HM-16.6 in the LDP and LD configuration. 
The multi-resolution DNN~\cite{lee2020deep} is individually trained for interpolation and extrapolation, then integrated with HM-16.6. Therefore, their coding gain is compared with our method over all three common test conditions.

The table shows luminance BD-Bitrate of each method against its corresponding version of HM. The network size is shown in the top row. Note that~\cite{extrapolation_with_reference_alignment} requires a separate extrapolation network for each QP value. Although the size of each such network is small (0.58M parameters), the total size grows with the number of QP values of interest. 
In practice, this is likely to be higher, because a codec with only four valid QP values seems too restrictive. For example, rate control algorithms~\cite{URQ_JSTSP_2013} often need to use a wide range of QP values in order to satisfy various rate constraints. In order to support all 52 QP values in HEVC, the approach in~\cite{extrapolation_with_reference_alignment} would require 52$\times$0.58=30.16M parameters, much more than our model. Even with only 10 supported QP values, this approach requires 10$\times$0.58=5.8M parameters, which is higher than our model. 
Nonetheless, under the common test conditions with four QPs, the total size of networks from~\cite{extrapolation_with_reference_alignment} 
is about 2.4 times smaller than our model. Meanwhile, the DNN in~\cite{lee2020deep} is very large, with about 43.36M parameters,\footnote{The size of the network was not reported in that paper. We did, however, re-produce their DNN architecture and from there estimated the size of the network.} due to the cascaded form of two FRUC networks~\cite{Niklaus_ICCV_2017}. So it is about 8 times bigger than our model.

Compared with~\cite{extrapolation_with_reference_alignment}, our model shows improved coding performance of up to 1.8\% in Class D for LDP and about 1.3\% in Classes C, D, and E for LD. Moreover, our method has the added benefit of being generic enough to operate over a range of QP values, and able to handle both extrapolation and interpolation. However, our gains relative to~\cite{extrapolation_with_reference_alignment} are smaller on Class B sequences, which have the highest resolution in Table~\ref{tbl:ldp_ld_work}. We believe this is due to the fact that the models in~\cite{extrapolation_with_reference_alignment} were trained on image patches from high-resolution (4K) sequences, and might be better suited for these types of sequences. Meanwhile,~\cite{lee2020deep} achieves higher coding gain than our model in LD configuration, while our model is better than~\cite{lee2020deep} in RA configuration. In the LDP configuration,~\cite{lee2020deep} and our model achieve the same average coding gain against HM-16.6. Similar to our integration method,~\cite{lee2020deep} also replaces an existing frame in a reference list with their DNN-generated frame. However, their model is considerably larger, and trained on higher-resolution sequences. Nonetheless, in spite of much smaller size, our model achieves competitive performance with~\cite{lee2020deep}, and also offers higher intepretability in the sense of providing affine motion parameters as a by-product of frame prediction.  


\section{Conclusion}
\label{sec:conclusion}
We presented a deep neural network (DNN) aimed at supplementing the conventional motion estimation and compensation for improved video coding efficiency. The proposed DNN includes a trainable affine estimation module to estimate motion between input frames and the target frame, and subsequently applies content-adaptive filters, generated by another module, to synthesize the target frame. This strategy, along with the use of dilated convolutions, makes our model smaller than most of the existing approaches in the literature, while achieving better coding efficiency. 
The proposed DNN was trained using a custom loss function, which includes a DCT-based $\ell_1$-loss term with several transform sizes, a multi-scale MSE loss term, and the object feature reconstruction loss term. While the same DNN architecture supports both uni- and bi-directional prediction, we trained two separate sets of weights for these two cases, and incorporated the resulting models into HEVC using two different integration approaches. 
In terms of coding efficiency, significant bit savings compared to HEVC were demonstrated. Furthermore, compared to the recent works on DNN-based frame prediction for video coding, the proposed methods also show improved coding performance while using a smaller model compared to most previous approaches. 


%




\ifCLASSOPTIONcaptionsoff
  \newpage
\fi



%


\bibliographystyle{IEEEtran}
\bibliography{ref}

\end{document}